\documentclass[prb,twocolumn,superscriptaddress]{revtex4-1}

\usepackage{graphicx}
\usepackage{hyperref}
\usepackage{dcolumn}
\newcolumntype{d}{D{.}{.}{2.5}}

\begin{document}


\title{First-principles investigation of competing magnetic interactions in (Mn,Fe)Ru$_2$Sn Heusler solid solutions}



\author{Elizabeth Decolvenaere}
\affiliation{Chemical Engineering Department, University of California, Santa Barbara, California 93106, USA}

\author{Michael Gordon}
\affiliation{Chemical Engineering Department, University of California, Santa Barbara, California 93106, USA}
\affiliation{Materials Research Laboratory, University of California, Santa Barbara, California 93106, USA}

 \author{Ram Seshadri}
\affiliation{Materials Department, University of California, Santa Barbara, California 93106, USA}
\affiliation{Materials Research Laboratory, University of California, Santa Barbara, California 93106, USA}
\affiliation{Department of Chemistry and Biochemistry, University of California, Santa Barbara, California 93106, USA}

\author{Anton Van der Ven}
\email[]{avdv@engineering.ucsb.edu}
\affiliation{Materials Department, University of California, Santa Barbara, California 93106, USA}


\date{\today}

\begin{abstract}
    Many Heusler compounds possess magnetic properties well-suited for applications as spintronic materials. The pseudo-binary Mn$_{0.5}$Fe$_{0.5}$Ru$_2$Sn, formed as a solid solution of two full Heuslers, has recently been shown to exhibit exchange hardening suggestive of two magnetic phases, despite existing as a \textit{single} chemical phase. We have performed a first-principles study of the chemical and magnetic degrees of freedom in the Mn$_{1-x}$Fe$_{x}$Ru$_2$Sn pseudo-binary to determine the origin of the unique magnetic behavior responsible for exchange hardening within a single phase. We find a transition from antiferromagnetic (AFM) to ferromagnetic (FM) behavior upon replacement of Mn with Fe, consistent with experimental results. The lowest energy orderings in Mn$_{1-x}$Fe$_{x}$Ru$_2$Sn consist of chemically- and magnetically-uniform (111) planes, with Fe-rich regions preferring FM ordering and Mn-rich regions preferring AFM ordering, independent of the overall composition. Analysis of the electronic structure suggests that the magnetic behavior of this alloy arises from a competition between AFM-favoring Sn-mediated superexchange and FM-favoring RKKY exchange mediated by spin-polarized conduction electrons. Changes in valency upon replacement of Mn with Fe shifts the balance from superexchange-dominated interactions to RKKY-dominated interactions.

\end{abstract}


\maketitle 


\section{Introduction}
The unique electronic properties of full Heusler compounds, defined as L2$_1$-ordered XY$_2$Z alloys with X and Y as transition metals and Z as a main-group element, make them promising materials for applications in spintronics\cite{Graf2011}, superconductors\cite{Graf2011a}, magnetocalorics\cite{Planes2009}, and shape-memory devices\cite{Yin2015}. The flexibility of Heusler compounds derives from their ability to realize large and tunable changes in properties with small changes in the valence electron count\cite{Stearns1980,Kubler1983,Kurtulus2005,Kurtulus2006}. Different choices of X and Y can produce various magnetic properties, as both Z-mediated indirect exchange between second nearest neighbor pairs of X, and direct exchange involving X-Y interactions, can give rise to ferro-(FM), ferri-(FrM), or antiferro-(AFM) magnetic configurations\cite{Graf2011,Graf2011a}. Many Heuslers have been demonstrated to follow Slater-Pauling behavior\cite{Galanakis2002b, Kurtulus2005, Galanakis2006}, a subset of which are calculated to be half-metals\cite{Ishida1995a,Galanakis2002b,Sasoglu2005,Galanakis2006}, yielding a family of materials with unparalleled magnetic flexibility.

The (Mn,Fe)Ru$_2$(Ge,Sn) genus of the Heusler family neatly demonstrates the transition from AFM to FrM to FM behavior with a change in the ratio of Fe to Mn as the X element\cite{Mizusaki2011,Douglas2016}. Mizusaki et al.\cite{Mizusaki2011} and Douglas et al.\cite{Douglas2016} have shown that both the Mn$_{x}$Fe$_{1-x}$Ru$_2$Ge and Mn$_{x}$Fe$_{1-x}$Ru$_2$Sn alloys exhibit magnetic behavior indicative of two-phase coexistence, evidenced by an increase in magnetic coercivity, in spite of the fact that the alloys form a \textit{single} chemical phase, i.e., a solid solution on the X = Mn,Fe sublattice. The spike in coercivity is theorized to be a consequence of local AFM/FM domains and interactions between the two, creating exchange hardening (broadening of the hysteresis loop) as observed on the macroscopic scale. Exchange bias (shifting of the hysteresis loop) and exchange hardening are both well-studied phenomena in nanocomposites and nanostructured thin films\cite{Pebley2016,Pebley2017} containing distinct FM-favoring and AFM-favoring chemical phases. However, exchange hardening \textit{without} the appearance of a second chemical phase is a phenomenon that, at present, is rarely observed and poorly understood. While previous studies have performed limited \textit{ab-initio} calculations to probe the magnetic behavior of the MnRu$_2$Sn and FeRu$_2$Sn end-members\cite{Ishida1995a, Douglas2016}, no studies at intermediate compositions of the Mn$_{x}$Fe$_{1-x}$Ru$_2$Sn alloy have yet been performed.


Here, we perform density functional theory (DFT)\cite{Hohenberg1964,Kohn1965} calculations on the Mn$_{1-x}$Fe$_x$Ru$_2$Sn alloy across the entire composition range $x = [0, 1]$, and explore both chemical and magnetic degrees-of-freedom.
We first probe the origin of FM and AFM ordering in the FeRu$_2$Sn and MnRu$_2$Sn Heuslers, respectively, and seek to answer \textit{why} a small change in the electronic configuration causes a dramatic change in magnetic ordering.
We consider two models of magnetic interaction proposed for Heuslers to analyze our results: (a) indirect exchange, as a competition between AFM-favoring, Sn-mediated superexchange and FM-favoring, Ruderman-Kittel-Kasuya-Yoshida (RKKY) interactions between the X-sites containing Fe and Mn, and (b) direct exchange, as a consequences of (anti-,non-)bonding interactions between Ru and Mn/Fe. As MnRu$_2$Sn and Mn-rich Mn$_{x}$Fe$_{1-x}$Ru$_2$Sn alloys are known to be AFM or FrM, we next evaluate a large number of magnetic orderings on the Mn/Fe sub-lattice site, and study the interplay between magnetic and chemical stability. We predict the existence of a miscibility gap at 0K with Mn$_{1-x}$Fe$_x$Ru$_2$Sn separating into a two-phase mixture of pure Heuslers (Mn,Fe)Ru$_2$Sn for all $x$. The lowest-energy chemical configurations of Mn$_{x}$Fe$_{1-x}$Ru$_2$Sn at intermediate concentrations ($0 < x < 1$) offer insights into the low-temperature magnetic behavior of quenched solid solutions.

\section{Prior Work: Magnetism in Heuslers}

The discovery of magnetism in MnCu$_2$Sn by Heusler\cite{Heusler1903} was at the time unexpected, owing to the non-magnetic\footnote{Antiferromagnetism would not be proposed as a phenomena for another 30 years\cite{Neel1936}} behavior of the constituent elements. While magnetic ordering is not unexpected in MnRu$_2$Sn or FeRu$_2$Sn, as we (now) know both Mn and Fe to exhibit magnetic ordering phenomena in their pure elemental forms, we expect the physics driving magnetic ordering to be different in the (Mn,Fe)Ru$_2$Sn Heuslers than in pure Mn or Fe. Several key differences exist between the pure elements and their Heuslers: the presence of additional alloying elements in significant concentrations, the interatomic distance between (next)nearest Mn or Fe neighbors, and the overall crystal structure both of the Mn/Fe sublattice, and of the entire crystal. Understanding how and why magnetism persists in MnRu$_2$Sn and FeRu$_2$Sn, and how those magnetic interactions may be influenced by alloying Mn and Fe on the same sublattice, requires understanding how and why magnetic behavior is arises in Heuslers in general. 

Numerous in-depth studies on Heusler families have been published exploring the origin of magnetism and half-metallic behavior\cite{Galanakis2002,Galanakis2005a,Sasoglu2005}, the prevalence and consequences of Slater-Pauling behavior\cite{Galanakis2002b,Kurtulus2005,Galanakis2006}, and the FM or AFM coupling of magnetic moments\cite{Kubler1983}. These studies can be divided into two general categories: (a) where the Y transition metal is assumed to determine the lattice constant and the valence electron count, but does not participate in the X-X magnetic coupling\cite{Stearns1980,Williams1983,Kubler1983,Sasoglu2008}, and (b) where the X-Y interactions are assumed to \textit{dominate} the magnetic Hamiltonian\cite{Kurtulus2005,Sasoglu2004,Sasoglu2005a, Sasoglu2005,Kurtulus2006}. Both perspectives make similar predictions about the magnetic behavior of pure MnRu$_2$Sn or FeRu$_2$Sn, but yield different insights into microscale magnetic ordering phenomena possibly present in (Mn,Fe)Ru$_2$Sn alloys of intermediate composition. To best understand which mechanisms determine the AFM-to-FM transition in Mn$_{1-x}$Fe$_x$Ru$_2$Sn with increasing $x$, we examine both models and apply the analysis methods used in each to the pure Heuslers (Mn,Fe)Ru$_2$Sn.


\subsection{Slater-Pauling Behavior}
Galanakis et al.\cite{Galanakis2002b,Galanakis2006} have demonstrated that many full Heusler alloys follow Slater-Pauling behavior, such that the total ferromagnetic moment is a linear function of the valence electrons:
\begin{equation}
    M_t = Z_t - 24,\label{one}
\end{equation}
where $M_t$ is the total moment, $Z_t$ is the number of valence electrons, and 24 is the total number of occupied spin-up plus spin-down valence bands in a traditional XY$_2$Z Heusler. 
The number 24 arises from the 12 electrons occupying the minority bands, which require 12 electrons in the majority bands to reach zero moment. Additional valence electrons do not change the occupation of the minority states, such that the moment is determined by the number of electrons past parity\footnote{at roughly 30 electrons, Galanakis et al. note that this rule begins to break down: the exchange-splitting required to push electrons into the highest-energy anti-bonding orbitals is infeasibly large}. 
In the case of Mn- or FeRu$_2$Sn, the 12 occupied minority states are: one Sn \textit{s}, three Sn \textit{p}, five hybridized Fe/Mn-Ru$_2$ \textit{d} and another three non-bonding Ru$_2$ \textit{d} states\cite{Galanakis2002b}. 
Owing to exchange splitting, those 12 states, plus an additional three (Mn) or four (Fe) states, are occupied by majority spin electrons. 
The states closest to the Fermi level, being delocalized conduction electrons, are however not easily assigned to site-specific orbitals.

\subsection{Indirect-Only Model: Superexchange and RKKY Exchange}
The first model of exchange proposed for Heusler alloys describes a collection of localized moments on the X site (Mn or Fe here) interacting indirectly via competing AFM-favoring superexchange and FM-favoring RKKY couplings\cite{Stearns1980,Williams1983,Kubler1983}.
As the nearest-neighbor (NN) X-X distance in Heuslers is always larger than 2$r_X$, where $r_X$ is the Van der Waals radius of X, the majority of X-X exchange is presumed to be indirect. 
Superexchange interactions between next-nearest-neighbor (NNN) X-X pairs, favoring AFM orderings, arise from overlap between partially-occupied X \textit{d} orbitals ({e.g.} Mn 3\textit{d}) and neighboring fully-occupied Z \textit{p} orbitals ({e.g.}, Sn 5\textit{p}).
In contrast, RKKY exchange interactions between NNN X-X pairs can favor either FM or AFM couplings, and arise from the large localized moment on X inducing a decaying oscillation in the spin-polarization of the conduction electrons.
For the interatomic distances of relevance in the Mn$_{1-x}$Fe$_x$Ru$_2$Sn alloys, the dominating terms of RKKY interaction are FM-favoring\cite{Stearns1980,Shi1994}.


Both superexchange and RKKY exchange parameters scale inversely with the energy required to promote an electron from a X \textit{d}-state to the Fermi level and depend on the number and distribution of states both immediately below and above the Fermi level\cite{Shi1994,Sasoglu2008}. 
The scaling of the competing exchange parameters can be qualitatively understood using a perturbative approach, where the coupling constants $j_{RKKY}$ and $j_s$ at the gamma point in reciprocal space ($q \to 0$) reduce to\cite{Shi1994}:

\begin{equation}
    j_{RKKY}(0) = V^4 D(\epsilon_F) / E_h^2,\label{jrkky}
\end{equation}

\begin{equation}
    j_{s}(0) = V^4 \sum_{nk}^{\epsilon_{nk} > \epsilon_F} {(\epsilon_F - \epsilon_{nk} - E_h)}^{-3}.\label{js}
\end{equation}
$V$ is an electronic mixing parameter, $D(\epsilon_F)$ is the density of states at the Fermi level, $\epsilon_F$ is the Fermi energy, $\epsilon_{nk}$ is the energy of a state at k-point $k$ in band $n$, and $E_h$ is the energy required to promote an electron from a \textit{d}-state of X to the Fermi level. \c{S}a\c{s}{\i}o\u{g}lu et al.\cite{Sasoglu2008} performed an in-depth exploration of the competition between $j_{RKKY}$ and $j_s$, using Mn-based Heuslers as examples. Their general conclusion is that a large number of states at (or just below) the Fermi level favor RKKY-type exchange, while a large number of states just above the Fermi level favor superexchange. The relative stability between FM and AFM behavior can, therefore, be altered by either varying the number of electrons with the chemistry on the X site, or by changing the lattice constant (thereby changing the orbital overlap in superexchange, or the point chosen along the sign oscillation of RKKY exchange).

\subsection{Direct Model: Dominating X-Y Exchange}

The second model of exchange for Heuslers assumes that the Y sites actively participate in exchange. The effects of Y-mediated exchange are explored in-depth by \c{S}a\c{s}{\i}o\u{g}lu et al.\cite{Sasoglu2004,Sasoglu2005,Sasoglu2005a} in numerous Heusler systems for lighter Y elements such as Co and Fe.
Dronskowski et al.\cite{Kurtulus2006} offer an (anti-)bonding-motivated discussion of the origin of AFM/FM behavior in Heuslers, demonstrating that even in cases where the Y element is a heavier element ({e.g.}, Ru), X-Y exchange interactions play a leading role in determining the magnetic configuration of the Heusler.
In cases where Y participates directly in exchange, two types of interactions must be considered for XY$_2$Z Heuslers: X-Y exchange and Y-Y exchange. As the two Y sites occupy interpenetrating FCC sub-lattices, the Y-Y distance is usually below 2$r_Y$, such that direct exchange between the two Y may be significant. The X-Y distance can also be smaller than $r_X+r_Y$, leading to direct X-Y exchange. This mechanism differs from the X-Z superexchange via Z's fully-occupied \textit{p} states, as both X and Y have partially-occupied \textit{d}-states ({e.g.}, Mn-Ru direct exchange would look different than Mn-Sn-Mn superexchange in MnRu$_2$Sn).

The potential Y-Y interactions resulting from direct exchange can be FM or AFM depending on the ratio of interatomic distance to the \textit{d}-orbital radius, in a similar fashion to pure transition metal elements on the Bethe-Slater curve\cite{Azumi1954}. Similarly, the X-Y interactions can be FM or AFM, while the (NN) X-X interactions are presumed to be purely FM owing to the large interatomic separation. The magnitudes of the Y-Y interactions are typically very small compared to those of the X-Y interactions or even the X-X interactions\cite{Kurtulus2005,Sasoglu2005}, thereby allowing the Y-Y interactions to be neglected. 



\section{Methods}
\subsection{\textit{Ab-Initio} Calculations}
All calculations were performed with VASP\cite{Kresse1994,Kresse1996a,Kresse1996} (version 5.4.1) using projector-augmented wave\cite{Blochl1994a,Kresse1999} (PAW) pseudo-potentials\footnote{dataset v.54, specific PAWs for each element were chosen using the guidelines at \url{https://cms.mpi.univie.ac.at/vasp/vasp/Recommended_GW_PAW_potentials_vasp_5_2.html}}. The generalized gradient approximation of Perdew, Burke, and Ernzerhof (PBE) was used for the exchange energy\cite{Perdew1996} and the interpolation formula of Vosko, Wilk, and Nusair was used for the correlation energy\cite{Vosko1980}. All calculations were performed spin polarized unless otherwise noted, with an energy cutoff of 540 eV, a $\Gamma$-centered Monkhorst-Pack k-point grid\cite{Monkhorst1976} with $17\times17\times17$ divisions in the unit cell (and scaled with reciprocal supercell size), and either the first order Methfessel-Paxton method\cite{Methfessel1989} (for relaxations) or the tetrahedron method with Bl\"{o}chl corrections\cite{Blochl1994} (for static runs) to integrate over electronic energy levels. Electronic relaxations were terminated with energy differences of less than $10^{-6}$ eV, and ionic relaxations were terminated by forces smaller than $10^{-2}$ eV/\AA\@.

All local properties ({e.g.}, site-specific magnetic moments or atomic charges) were determined using the Wigner Seitz radii provided with the pseudopotentials. Non-neutral calculations were performed utilizing a homogeneous compensating background charge, fixing lattice vectors and atomic coordinates to those of the lattice of the neutral parent structure.
Fixed total-moment calculations were performed by constraining the total magnetic moment, but allowing the per-site moments to relax from an initial FM or AFM configuration scaled from the FM or AFM moments achieved without total-moment constraints.
The LOBSTER software package\cite{Deringer2011,Maintz2013,Maintz2016a} (version 2.1.0) was used for the calculation of Crystal Orbital Hamilton Populations (COHP) to analyze bonding\cite{Dronskowski1993}. We utilized the pbeVaspFit2015 basis set\cite{Maintz2016} and selected per-element basis functions matching the electrons treated as valence electrons in the pseudopotentials.



\begin{table*}[t]
    \caption{Total and per-site magnetic moments for MnRu$_2$Sn and FeRu$_2$Sn obtained from DFT calculations using different lattice constants, as well as experimental measurements.\label{magtable}}
    \begin{ruledtabular}
    \begin{tabular}{c d d d d d d}
        \multicolumn{1}{c}{System} & \multicolumn{3}{c}{MnRu$_2$Sn Moments ($\mu_B$)} & \multicolumn{3}{c}{FeRu$_2$Sn Moments ($\mu_B$)} \\ \hline
        \multicolumn{1}{c}{} & \multicolumn{1}{c}{Total} & \multicolumn{1}{c}{Mn} & \multicolumn{1}{c}{Ru} & \multicolumn{1}{c}{Total} & \multicolumn{1}{c}{Fe} & \multicolumn{1}{c}{Ru}\\ \hline
            DFT (Full Relax)        & 0.0 & 3.155 & 0.078 & 4.141 & 3.065 & 0.497 \\
            DFT (Exp. Lat. Const.)  & 0.0 & 3.140 & 0.078 & 4.136 & 3.057 & 0.498 \\
            DFT (Mean Lat. Const.)  & 0.0 & 3.134 & 0.071 & 4.137 & 3.059 & 0.498 \\
            Experiment              & 0.0 & 3.4(1) & 0.0 & 3.4 & 3.16(4) & 0.4(1)
    \end{tabular}
    \end{ruledtabular}
\end{table*}

\subsection{Enumeration of Configurations}
The CASM software package\cite{Thomas2013,Puchala2016,Puchala2013,VanderVen2010} was used to enumerate a large number of symmetrically distinct chemical and magnetic orderings. The energies of all symmetrically-distinct configurations of chemical identity (Fe, Mn) and magnetic spin (negative or positive per-site moment) within supercells containing up to four unit cells were calculated using VASP with the settings described above. One configuration containing six unit cells at $x_{Fe} = 0.33$ was also considered, as the lowest-energy three-unit-cell supercell corresponds to a high-energy, magnetically-frustrated state.Volume and site relaxations were minor ($0.998 < V/V_0 < 1.009$, and mean-squared-displacement $< 0.7 \times 10^{-3}$ \AA); however, magnetic relaxations were significant. While VASP prohibits changes of the magnetic moments that would lower symmetry ({e.g.}, FM to AFM), changes that add additional symmetries, or move to a different magnetic configuration with the same symmetry, are allowed. Multiple instances of AFM structures relaxing to otherwise-enumerated FM structures were found; the initial AFM structures were considered unstable and were removed from further calculations. More interestingly, several cases were found where one AFM or FrM configuration would relax into its spin-reversed twin ({i.e.}, applying the time-reversal operator), and visa-versa ($A \to B$ and $B \to A$).
In the cases where multiple initial magnetic configurations relaxed onto the same final configuration, the structure with the lowest energy was kept and the others discarded.


\section{Results and Discussion}

\subsection{Electronic Structure of MnRu$_2$Sn and FeRu$_2$Sn}\label{pure}

To understand the potential causes of complex magnetic behavior in $0 < x < 1$ solid solutions of Mn$_{1-x}$Fe$_x$Ru$_2$Sn, we first sought to understand the difference in electronic structure between the $x=0$ and $x=1$ end-members. MnRu$_2$Sn possesses L1$_1$ AFM ordering ({i.e.}, (111) spin-up planes alternated by spin-down planes) on the Mn FCC sublattice with a large moment on Mn and no moment on Ru. FeRu$_2$Sn, in contrast, is FM with a large moment on Fe and a small-but-nonzero moment on Ru\cite{Douglas2016}. We examined many different AFM and FrM magnetic orderings of the pure Heuslers (Mn,Fe)Ru$_2$Sn. The experimentally-observed magnetic orderings (AFM for Mn and FM for Fe) were found to be the ground states, with energy differences of 24 meV and 87 meV (per primitive cell) to the next-lowest-energy magnetic orderings for MnRu$_2$Sn and FeRu$_2$Sn, respectively. In addition to the fully-relaxed structures, we calculated total and local moments of cells having the experimental lattice constants and of cells having an averaged lattice constant between the end-member constants. In both cases the internal ionic degrees of freedom were fully relaxed. Table~\ref{magtable} compares calculated total and local site magnetic moments with experimental total saturation magnetizations measured at 5T and 4K and per-site moments extracted from Rietveld refinements of neutron powder diffraction data.

\begin{figure}[b]
    \includegraphics[width=0.49\textwidth]{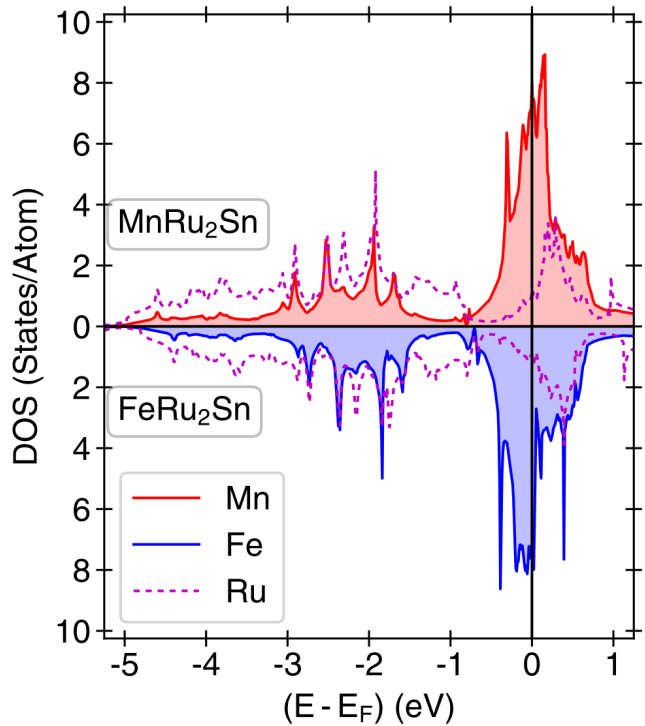}%
    \caption{Non-spin-polarized site-projected DOS for MnRu$_2$Sn (above) and FeRu$_2$Sn (below) for the Mn or Fe and Ru sites; Sn sites contribute states far below the Fermi level ($\leq -6$ eV) and are not shown. While the states far below the Fermi level are largely unaffected by the addition/subtraction of an extra electron/proton pair, the behavior at the Fermi level changes substantially.\label{nm_dos}}%
\end{figure}

The moments obtained from DFT in all cases compare favorably to the experimental results, except for the saturation magnetization in FeRu$_2$Sn, where the lower-than-anticipated experimental moment in the FM case arises from the dispersion of grain orientations and the presence of grain and magnetic domain boundaries in the experimental sample\cite{Douglas2016}. The relaxed DFT lattice parameters of 6.22 \AA{} and 6.21 \AA{} for MnRu$_2$Sn and FeRu$_2$Sn, are also in good agreement with the experimental lattice constants of  6.20 \AA{} and 6.19 \AA{} as measured with neutron diffraction at 15K by Douglas et al.\cite{Douglas2016}. The fully-relaxed lattice constants are slightly larger than the experimental values; the lattice constant is set by Ru (the element with the largest radius), and PBE is known to systemically overestimate lattice constants for 5\textit{d} elements\cite{Schimka2013}.



We can understand the transition from AFM to FM upon replacement of Mn by Fe in (Mn,Fe)Ru$_2$Sn, as well as the likely intermediate states for a disordered Mn/Fe solid solution, by analyzing how the electronic structure around the Fermi level changes with a change in the X element. To most easily compare reciprocal-space properties, all of the following calculations were performed using the mean lattice constant between the experimental MnRu$_2$Sn and FeRu$_2$Sn values. The density of states (DOS) of the non-spin-polarized (referred to hereafter as non-magnetic, or NM, for simplicity\footnote{Non-spin-polarized should not be taken to mean paramagnetic, in this context. While non-spin-polarized calculations have sometimes been used as a proxy for paramagnetic configurations, we stress that this approach is generally incorrect.}) Heuslers are shown in Figure~\ref{nm_dos}, along with the projected DOS of the Mn/Fe and Ru \textit{d}-states.
Figure~\ref{nm_dos} clearly shows large changes in the states available above and below the Fermi level when going from NM MnRu$_2$Sn to NM FeRu$_2$Sn. In the Mn Heusler, there are more states just above the Fermi level than below it, while in the Fe Heusler, the reverse is true. Far below the Fermi level the electronic structures of both Heuslers are nearly the same.
At the Fermi level, the difference between Mn and Fe resembles that of a rigid-band model, where the addition of an additional electron rigidly shifts the Fermi level upward to accommodate one more state.

We can connect these differences in the Fermi-level states back to the indirect-exchange-only model of magnetic ordering, {i.e.}, as a competition between superexchange and RKKY exchange. Both Mn- and Fe-rich Heuslers have a large DOS at the Fermi level, driving a large RKKY term, but in Fe there is a large drop-off just past the Fermi level. The superexchange interaction, $j_s$ in Equation~\ref{js}, is bounded from above by $j_s(0) \leq V^4 N/E_h^3$, where N is the number of unoccupied states near the Fermi level\cite{Shi1994}. Therefore, in Mn, where N is large, $|j_s|$ should be greater than $|j_{RKKY}|$, and AFM behavior should dominate. In Fe, where N is smaller, $|j_s|$ should be smaller than $|j_{RKKY}|$, and FM behavior should dominate. It is worth noting here that $D(\epsilon_F)$ in the Fe-Heusler is also smaller than in the Mn-Heusler, and so the overall magnitudes of the coupling constants are expected to be smaller.

\begin{table}[t]
    \caption{Relative formation energies per primitive cell of various magnetic and electronic configurations of MnRu$_2$Sn and FeRu$_2$Sn.
    All energies represent fully-relaxed structures.\label{magetable}}
    \begin{ruledtabular}
    \begin{tabular}{c c c c}
        \multicolumn{1}{c}{System} & \multicolumn{1}{c}{FM - NM} & \multicolumn{1}{c}{AFM - NM} & \multicolumn{1}{c}{FM - AFM} \\ 
        \multicolumn{1}{c}{} & \multicolumn{1}{c}{(eV)} & \multicolumn{1}{c}{(eV)} & \multicolumn{1}{c}{(meV/$\mu_B$)} \\ 
            MnRu$_2$Sn               & -1.305 & -1.357 & 16.7  \\ 
            {[}MnRu$_2$Sn{]}$^{-1}$  & -2.025 & -2.036 & 2.6   \\ 
            FeRu$_2$Sn               & -1.084 & -0.998 & -20.8   \\ 
            {[}FeRu$_2$Sn{]}$^{+1}$  & -0.716 & -0.753 & 11.9   \\ 
    \end{tabular}
    \end{ruledtabular}
\end{table}

\begin{figure*}[t]
    \includegraphics[width=0.95\textwidth]{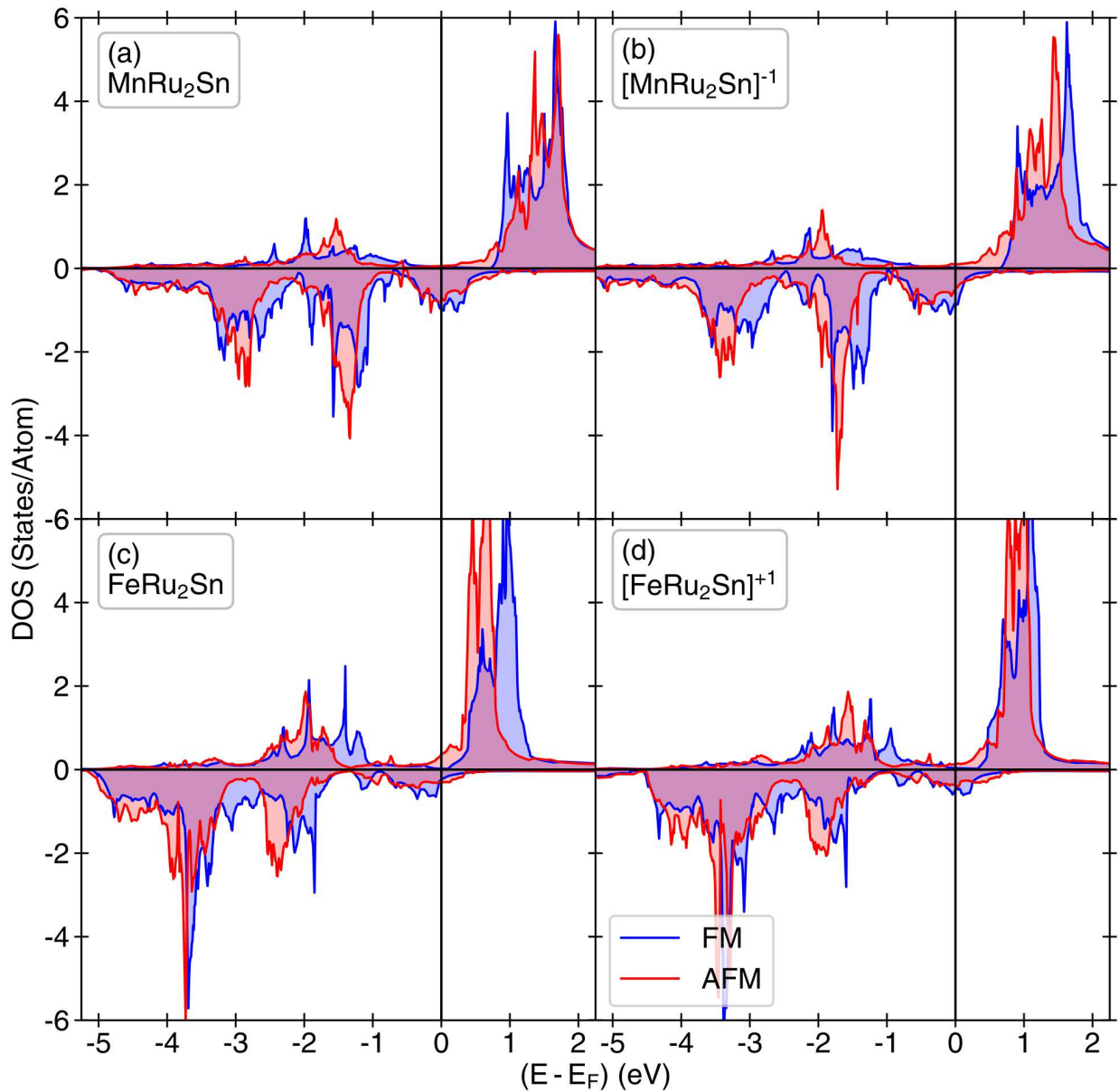}%
    \caption{Spin-polarized site-projected DOS for MnRu$_2$Sn (a,b) and FeRu$_2$Sn (c,d), showing the projected per-site Mn or Fe DOS for both FM and L1$_1$ AFM configurations. Panels (b) and (d) have had an electron added or subtracted, respectively, such that the Mn-Heusler has the same number of electrons as the Fe-Heusler, and vise-versa. The total DOS, while is not shown, has occupations at the Fermi level nearly identical to those of the Mn/Fe-projected DOS.\label{spin_dos}}%
\end{figure*}

The differences in the states near the Fermi level persist into the spin-polarized calculations. The Mn or Fe site-projected DOS for the FM and L1$_1$ AFM configurations are shown in Figures~\ref{spin_dos} a and c, with corresponding formation energies in Table~\ref{magetable}.

\begin{table*}[t!]
    \caption{Estimated coupling constants $j_i$ fit to the results of Figure~\ref{magcurves} using Equation~\ref{model}, and the magnitude of the terms for the lowest-energy ferromagnetic and antiferromagnetic configuration of each Heusler. 95\% confidence intervals are given for the $j_i$s, while the $R^2$ values are 0.987 and 0.996 for the Mn- and Fe-Heuslers, respectively.\label{jtable}}
    \begin{ruledtabular}
        \begin{tabular}{c c c c c c}
                System & \multicolumn{5}{c}{Coupling (meV/$\mu_B^2$)} \\
                       & $j_0$            & $j_1$            & $j_2$           & $j_3$          & $j_4$ \\
            MnRu$_2$Sn & $-353 \pm 104$   & $331 \pm 30$     & $-31.8 \pm 3.7$ & $33.4 \pm 3.5$ & $30.9 \pm 8.2$ \\
            FeRu$_2$Sn & $-85.1 \pm 66.6$ & $116.3 \pm 37.4$ & $-33.2 \pm 3.4$ & $32.1 \pm 2.9$ & $-15.9 \pm 12.3$ \\ \hline
              FM State & \multicolumn{5}{c}{Contribution (meV)} \\
                       &                  & $3j_1\langle \phi^{NN}_{Ru} \rangle$ & $6j_2\langle \phi^{NN}_{Mn/Fe} \rangle $ & $3j_3 \langle \phi^{NNN}_{Mn/Fe} \rangle$ & $4j_4 \langle \phi^{NN}_{Ru,Mn/Fe} \rangle$ \\
            MnRu$_2$Sn &                  & $1.8 \pm 0.2$    & $-1735 \pm 200$ & $911 \pm 95$   & $-15.6 \pm 4.2$  \\
            FeRu$_2$Sn &                  & $86.5 \pm 27.8$  & $-1864 \pm 193$ & $901 \pm 80$   & $97.1 \pm 75$ \\ \hline
             AFM State & \multicolumn{5}{c}{Contribution (meV)} \\
                       &                  & $3j_1\langle \phi^{NN}_{Ru} \rangle$ & $6j_2\langle \phi^{NN}_{Mn/Fe} \rangle $ & $3j_3 \langle \phi^{NNN}_{Mn/Fe} \rangle$ & $4j_4 \langle \phi^{NN}_{Ru,Mn/Fe} \rangle$ \\
            MnRu$_2$Sn &                  & $0$              & $0$             & $-984 \pm 102$ & $13.8 \pm 3.7$  \\
            FeRu$_2$Sn &                  & $0$              & $0$             & $-898 \pm 80$  & $24.8 \pm 19.2$
    \end{tabular}
    \end{ruledtabular}
\end{table*}

For the Mn-Heusler, the Fermi level resides at approximately the middle of a broad peak of both the projected DOS at the Mn site (see Figure~\ref{spin_dos} (a)) and the total DOS. This matches the behavior of the NM states, suggesting dominance of the superexchange interaction over the RKKY interaction, resulting in the energy of the AFM phase being more than a dozen meV (per primitive cell) below that of the FM phase. Both the FM and AFM orderings are also far lower in energy ($>$1 eV per primitive cell) than the NM state, indicating a large driving force towards magnetization.

The Fermi level in the FM Fe-Heusler lies at the start of a near-gap where the DOS is at a minimum for the minority spin, and just past a peak in the majority spin (see Figure~\ref{spin_dos} (c)). The presence of the majority-spin peak immediately before the Fermi level leads to a large RKKY contribution to the coupling in the FM configuration of the Fe-Heusler. In the AFM configuration, however, the very local maximum immediately after the Fermi level is still smaller than the maximum before the Fermi level in the FM configuration. This implies that the degree of superexchange (AFM-favoring) coupling in the AFM configuration is far smaller than the degree of RKKY (FM-favoring) coupling in the FM configuration, leading to FM being lower in energy.



To establish the importance of nuclear charge/effective potential on magnetism, we also considered FM and AFM configurations of Mn- and FeRu$_2$Sn when one electron is added or subtracted, respectively. The change in total electron count serves to give MnRu$_2$Sn the same valency as FeRu$_2$Sn, and vice-versa. When an additional electron is added to MnRu$_2$Sn, the Fermi level moves to higher energy, but the dispersion of the bands also broadens as the additional electron induces further delocalization, as seen in Figure~\ref{spin_dos} (b). As a consequence, while the energy difference between FM and AFM orderings decreases upon addition of an extra electron to MnRu$_2$Sn, there are still a sufficient number of states above the Fermi level to (barely) prefer the AFM ordering.

Conversely, when an electron is subtracted from FeRu$_2$Sn, the states become more localized (owing to less Coulomb repulsion) \textit{and} the Fermi level moves to lower energies, putting the Fermi energy directly in the middle of the large DOS peak in the FM configuration as seen in Figure~\ref{spin_dos} (d). This new distribution of states around the Fermi level implies that $|j_s|$ is now much closer in magnitude to $|j_{RKKY}|$, destabilizing the FM state.
In contrast, the removal of an electron from AFM FeRu$_2$Sn has virtually no effect on the electronic structure near the Fermi level. As a result, the energy change upon removal of an electron penalizes the FM state more than the AFM, so much so that for FeRu$_2$Sn it alters the ground state from FM to AFM\@. This effect becomes evident upon inspection of the change in energy when going from the NM state to the FM and AFM states. Table~\ref{magetable} shows that the AFM state gains more in energy compared to the FM state relative to the NM state in FeRu$_2$Sn upon subtraction of an electron (by approximately 100 meV per primitive cell).

\begin{figure}
    \includegraphics[width=0.49\textwidth]{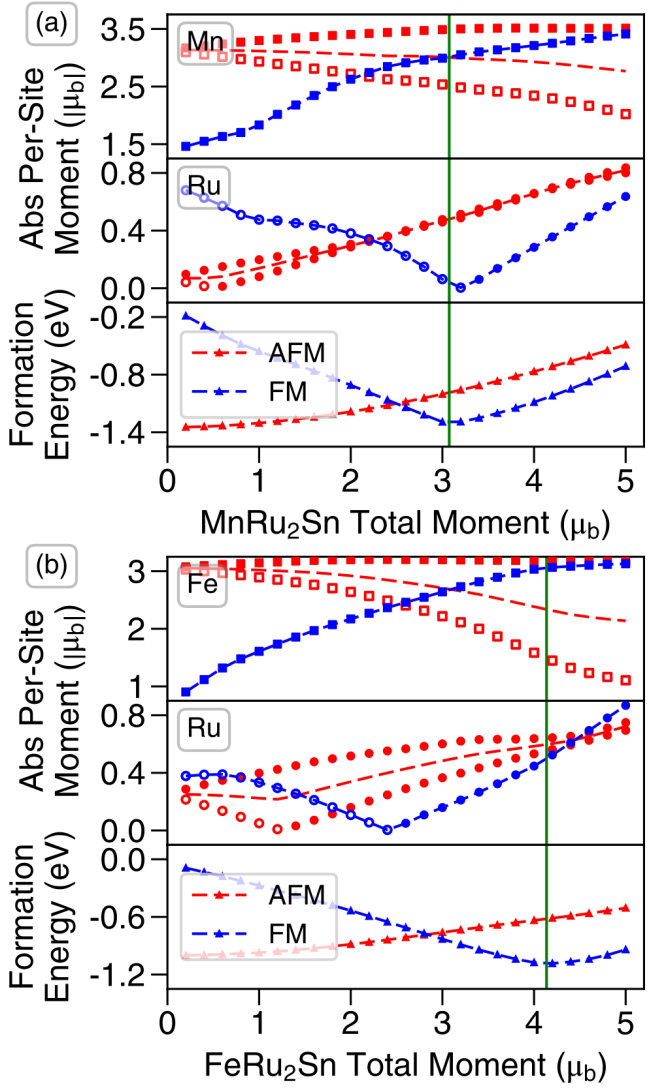}%
    \caption{Formation energies and site-specific magnetic moments for total-moment-constrained DFT calculations in MnRu$_2$Sn (a) and FeRu$_2$Sn (b). Absolute values of magnetic moments are shown in the top two plots of (a) and (b), where the chemical symbol specifies which site-projection is being measured, and full versus hollow symbols indicate a positive or negative sign of the moment, respectively. In the AFM cases, the moments of both the symmetrically-distinct Mn/Fe/Ru sites are indicated, while the average of the absolute value of the moment is given by the dashed line bisecting the sets of markers. Green vertical lines indicate the equilibrium (minimum-energy) total moment.\label{magcurves}}%
\end{figure}

There is another subtle, yet relevant difference between MnRu$_2$Sn and FeRu$_2$Sn: in the FM configuration, Mn and Ru are AFM aligned to one another, while Fe and Ru are FM to one another.
This suggests that there are also interactions that couple the X ({i.e.}, Mn and Fe) and Y ({i.e.}, Ru) sites, and that these couplings may be important.
We can estimate the relative strengths of Ru-Ru, Ru-Mn/Fe and Mn/Fe-Mn/Fe interactions by fitting a rudimentary magnetic Hamiltonian to DFT supercell calculations at fixed total moment. This calculation is in the spirit of the Heisenberg Hamiltonian which has been successfully applied to similar Heusler systems\cite{Sasoglu2005, Sasoglu2005a, Thoene2009, Grunebohm2016}. Our model Hamiltonian takes the form:
%
%
\begin{eqnarray}
    E &= j_0 + 3j_1 \langle \phi^{NN}_{Ru} \rangle + 6j_2 \langle \phi^{NN}_{Mn/Fe} \rangle \nonumber \\
    & \quad + 3j_3 \langle \phi^{NN}_{Mn/Fe} \rangle + 4j_4 \langle \phi^{NN}_{Ru,Mn/Fe} \rangle,\label{model}
\end{eqnarray}
where $E$ is the energy per primitive cell, $j_0$ can be interpreted as the energy gain from exchange splitting, $j_1$ is the Ru-Ru coupling, $j_2$ and $j_3$ are the nearest-neighbor and next-nearest-neighbor Mn/Fe-Mn/Fe couplings, respectively, and $j_4$ is the Ru-Mn/Fe coupling. The $\phi$ are the products of site-projected moments for Ru, Fe, or Mn, for either nearest-neighbor (NN) or next-nearest-neighbor (NNN) pairs. These are equivalent for dot products of magnetic vectors in a system where only $\theta = 0$ or $\theta=180$ are allowed. The coefficients 3, 6, 4, and 3 correspond to the number of (next-)nearest-neighbor pairs per primitive cell.
We stress that \textit{the $j_i$ are not the true exchange constants} and are not suitable for building a more complex model. However, the magnitudes of the $j_i$ should give an indication of the relative strength of each type of interatomic coupling. Curves of the formation energy (referenced against the NM calculation) versus the net moment as calculated with PBE are shown in Figures~\ref{magcurves} a and b, with the values of $j_0$, $j_1$, $j_2$, $j_3$, and $j_4$ fit to these curves given in Table~\ref{jtable}.

The leading terms in both MnRu$_2$Sn and FeRu$_2$Sn are the $j_0$ exchange-splitting energy.
While the $j_1$ terms have similar magnitudes as the $j_0$ terms, the \textit{contributions} from the Ru-Ru interactions are overall small, as the magnitude of the Ru moments are 0.1 to 0.01 times smaller than the Fe or Mn moments, respectively.
The calculated contributions of each term for the lowest-energy FM configuration of each Heusler are given in the second part of Table~\ref{jtable}, making the difference in magnitudes more obvious.
These results tell us three things: (a) Mn-Mn interactions dominate in MnRu$_2$Sn, (b) Fe-Fe interactions are the most important in FeRu$_2$Sn, and (c) Ru-Ru and Ru-Fe interactions are also relatively important in FeRu$_2$Sn.
From the signs of the $j_i$'s we can also conclude that Ru-Ru prefers an AFM-type ordering (though the energy change for this preference is small), and that Ru-Mn prefers an AFM-type ordering (again, small), while Ru-Fe prefers a FM-type ordering (larger).


\begin{figure}
    \includegraphics[width=0.49\textwidth]{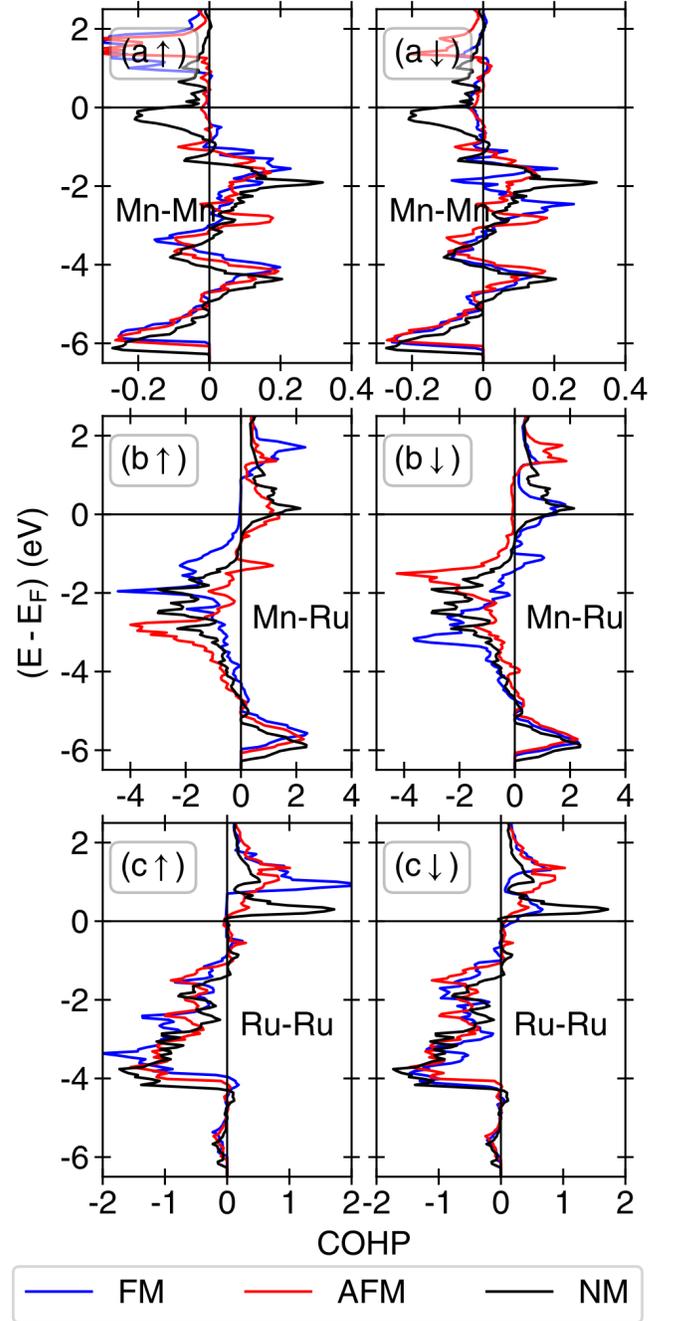}%
    \caption{COHP bonding analysis for various bonds present in MnRu$_2$Sn for NM, FM, and AFM configurations. For FM and AFM, the left column of plots represents the majority spin, while the right column of plots represents the minority spin.\label{COHP_mn}}%
\end{figure}

\begin{figure}
    \includegraphics[width=0.49\textwidth]{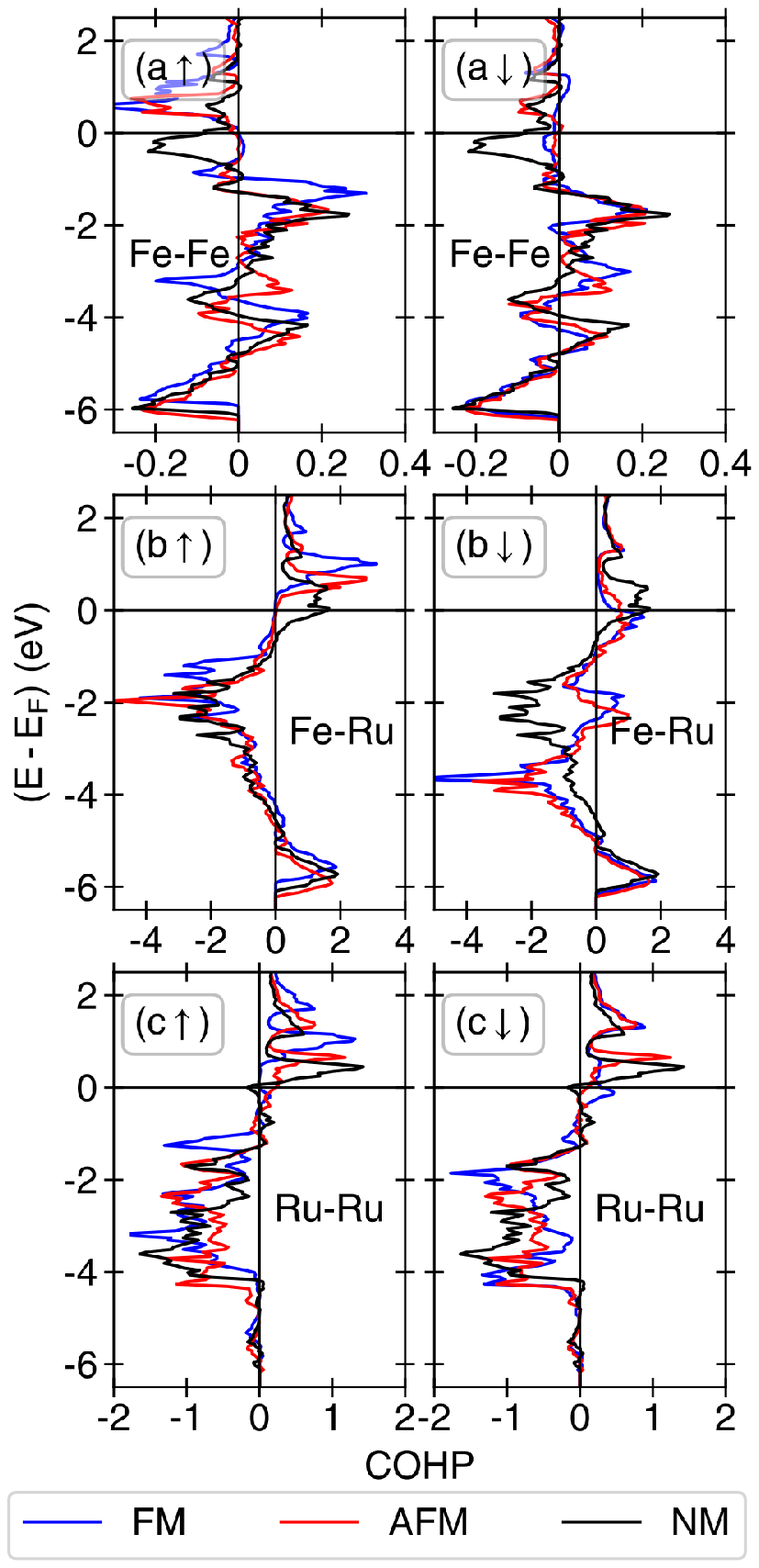}%
    \caption{COHP bonding analysis for various bonds present in FeRu$_2$Sn for NM, FM, and AFM configurations. For FM and AFM, the left column of plots represents the majority spin, while the right column of plots represents the minority spin.\label{COHP_fe}}%
\end{figure}

We can gain further insight about the relative importance of Ru-Ru, Mn/Fe-Mn/Fe, and Ru-Mn/Fe interactions with the help of a Crystal Orbital Hamiltonian Populations (COHP), bonding analysis\cite{Dronskowski1993} in the style of Kurtulus, et al.\cite{Kurtulus2006}. COHP provides an ``energy-resolved visualization of chemical bonding'' and enables the easy visualization of the bonding, anti-bonding, or non-bonding behavior between a pair of sites in a solid. The COHP procedure weights the electronic DOS by entries from the Hamiltonian matrix, {i.e.}, overlap of eigenstates (rather than of orbitals as done in a crystal orbital overlap populations analysis). The resulting set of COHP and energy values looks similar to a DOS plot, but convey different (but related) information: negative values correspond to bonding states, positive values correspond to anti-bonding states, and values near zero are interpreted as non-bonding states. The integrated COHP (up to the Fermi level) indicates the total (anti-,non-)bonding character of the interaction; the lowest-energy structure should be the one that maximizes bonding.

A more in-depth discussion of the COHP technique and how it applies to the analysis of magnetic materials (especially transition metals) is presented in a review paper by G. A. Landrum and R. Dronskowski\cite{Landrum2000}. We offer a brief summary. An alternate (but equivalent) view on the origin of itinerant magnetism can be developed by considering how spin-polarization can alleviate anti-bonding or non-bonding interactions otherwise present in a system without spin-polarization (referred to as nonmagnetic or NM). One can examine the COHP at the Fermi level of a material calculated in a NM framework and use those results to predict the type of exchange splitting and magnetic interactions (if any) of the material in a spin-polarized framework. Strongly anti-bonding interactions at the Fermi level in the NM calculation correspond (typically) to ferromagnetic interatomic interactions in the spin-polarized calculation, while non-bonding interactions yield antiferromagnetic interactions. For MnRh$_2$Ge (the most similar to (Mn,Fe)Ru$_2$Sn) Kurtulus, et al.\cite{Kurtulus2006} argue that the anti-bonding character of the Rh-Rh and Rh-Mn interactions in the NM calculation, and \textit{not} the Mn-Mn interactions, drive exchange splitting and FM ordering in the system. 


Figures~\ref{COHP_mn} and~\ref{COHP_fe} show COHP analyses for MnRu$_2$Sn and FeRu$_2$Sn, respectively, contrasting NM, FM, and AFM configurations. All COHP calculations were performed with the same supercell (containing four primitive cells) and having the mean experimental lattice constant described earlier. In contrast to results for the similar Heusler Rh$_2$MnGe\cite{Kurtulus2006}, we find only non-bonding interactions at the Fermi level for the Ru-Ru pair in MnRu$_2$Sn (with a strong anti-bonding peak \textit{above} the Fermi level). An anti-bonding peak is present at the Fermi level for the Mn-Ru pair in the NM state of MnRu$_2$Sn, which is not reduced significantly upon spin-polarization. These results are consistent with our estimates of the various $j_i$ in Table~\ref{jtable}, and exclude the possibility of significant Ru-Mn, Ru-Ru, or Mn-Mn (NN) interactions. By eliminating direct-exchange interactions from our model using COHP analysis, we strengthen our conclusion that NNN Mn-Mn interactions are the dominating force in determining the magnetic configuration of MnRu$_2$Sn.

In FeRu$_2$Sn, the COHP analysis (Figure~\ref{COHP_fe}) reveals that the Fe-Ru interaction is still anti-bonding at the Fermi level, though the magnitude of the COHP value in either the FM or AFM spin-polarized configurations in FeRu$_2$Sn is smaller than that of spin-polarized MnRu$_2$Sn. This difference is reflected in the difference between the $j_4$ contributions for the two Heuslers seen in Table~\ref{jtable}; $j_4\langle\phi^{NN}_{Ru,Fe}\rangle$ is about six times larger in FeRu$_2$Sn than in MnRu$_2$Sn. For the Ru-Ru pair, spin-polarization turns a bonding interaction into an \textit{anti-bonding} interaction, but the COHP magnitude in both cases is small. Neither the magnitude (or sign) of the Ru-Ru interactions, nor the magnitude of the Ru-Fe interaction, match the COHP values seen in MnRh$_2$Ge or MnCo$_2$Ga\cite{Kurtulus2006}.

Overall, our findings imply that magnetism in either the Mn- or the Fe-Heusler is primarily controlled by Mn/Fe-Mn/Fe interactions, with Ru-Fe interactions playing a much smaller secondary role in the Fe-Heusler. Without making a broader statement about all Heuslers, we conclude that a competition between FM-favoring RKKY and AFM-favoring Sn-mediated superexchange determines the magnetic configuration in (Mn/Fe)Ru$_2$Sn. As a consequence, a reasonable model of magnetism in the disordered Mn/Fe solid solution can be constructed utilizing only the Mn/Fe sublattice, without the need to consider a more complex model that explicitly includes the Ru sublattices. Instead, the Ru moment becomes a dependent variable following the neighboring Fe moments, as the Fe-Ru coupling is several orders of magnitude larger than for Mn-Ru. As the moment on the Sn sites is never larger than 0.001 $\mu_B$, we can also disregard Sn-Mn/Fe contributions. The role of Sn, instead, is to facilitate superexchange via next-nearest-neighboring Mn/Fe sites.

By simplifying to a model dependent only on the properties of the Mn/Fe sublattice, the remaining problem becomes somewhat straightforward: how do Mn and Fe prefer to organize in solution, does anything unexpected happen upon mixing Mn and Fe, and what happens to the \textit{d}-bands and the Fermi-level occupation in the range of intermediate compositions?

\subsection{Magnetic and chemical coupling in Mn$_{1-x}$Fe$_x$Ru$_2$Sn}

\begin{figure}[b]
    \includegraphics[width=0.49\textwidth]{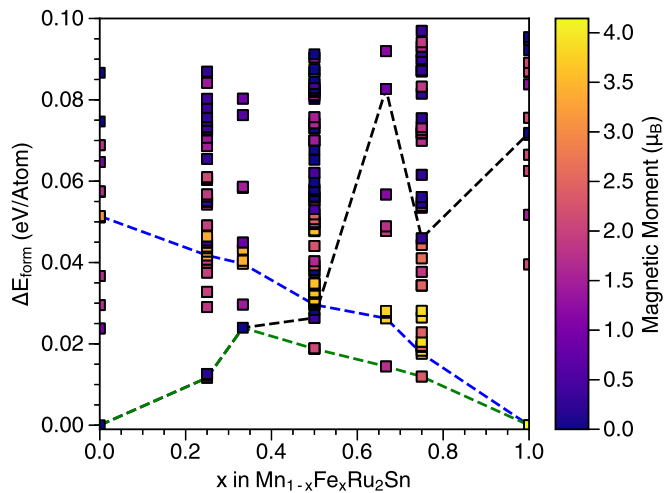}%
    \caption{Formation energies of the enumerated magnetic and chemical orderings in Mn$_{1-x}$Fe$_{x}$Ru$_2$Sn supercells, colored by the absolute value of the net magnetic moment. The dashed blue and black lines trace the lowest-energy FM ($\mu_B>3$) and AFM ($\mu_B < 1$) structures, respectively, while the dashed green line joins the overall lowest-energy structures.\label{hull}}
\end{figure}

We considered 193 symmetrically distinct chemical/magnetic configurations on the Mn/Fe sublattice of the full Heusler Mn$_{1-x}$Fe$_x$Ru$_2$Sn enumerated in supercells containing up to four copies of the primitive cell (16 total atoms).
The calculated formation energies of these structures, referenced against FM FeRu$_2$Sn and L1$_1$-AFM MnRu$_2$Sn, are shown in Figure~\ref{hull} (a). All formation energies at intermediate compositions $x$ between 0 and 1 are positive, indicating the existence of a chemical miscibility gap at low temperature. In the thermodynamic limit, Mn$_{1-x}$Fe$_x$Ru$_2$Sn alloys will, therefore, phase separate into regions that are Mn rich and regions that are Fe rich. At sufficiently high temperatures, a solid solution will become stable in which Mn and Fe are uniformly distributed over the X sublattice of Mn$_{1-x}$Fe$_x$Ru$_2$Sn, lacking any long-range order. When quenched from such a temperature ({e.g.}, 1173K as performed by Douglas et al.\cite{Douglas2016}), sluggish kinetics can be used to lock in the disordered solid solution.

The absence of long-range order, however, does not mean that the Mn and Fe lack any order at all. Most solid solutions exhibit a substantial degree of short-range order due to strong local energetic interactions that are not completely overwhelmed by entropy. The low energy configurations at intermediate compositions of Figure~\ref{hull} (a) can serve as useful structural models with which to analyze chemical and magnetic interactions in local environments that are representative of the high temperature solid solutions that have been annealed and then quenched. The lowest-energy intermediate-composition phases contain clustered domains of Mn-rich regions next to Fe-rich domains\footnote{It is worth noting that, had we enumerated larger supercells, the lowest-energy structures in the intermediate composition range would asymptotically approach zero. This is a result of the Mn/Fe miscibility gap: larger structures would allow larger and larger volumes of Mn-rich and Fe-rich domains, decreasing the surface-area-to-volume ratio of the interface.}.

In all the structures considered in this study, the per-site magnetic moments for the (Mn,Fe) sites relaxed to values that range from 2.7 to 3.3 $\mu_B$, with lower values at intermediate compositions and higher values for pure Heuslers (Mn,Fe)Ru$_2$Sn.
The magnetic moments on the Ru and Sn sites were universally small, $< 0.7 \mu_B$ for Ru and $< 0.022 \mu_B$ for Sn.
Trends for FM and AFM structures with the lowest energy are indicated with the dark dashed lines in Figure~\ref{hull} (a)  and show a cross-over at equiatomic composition.



\begin{figure*}[t!]
\includegraphics[width=0.95 \textwidth]{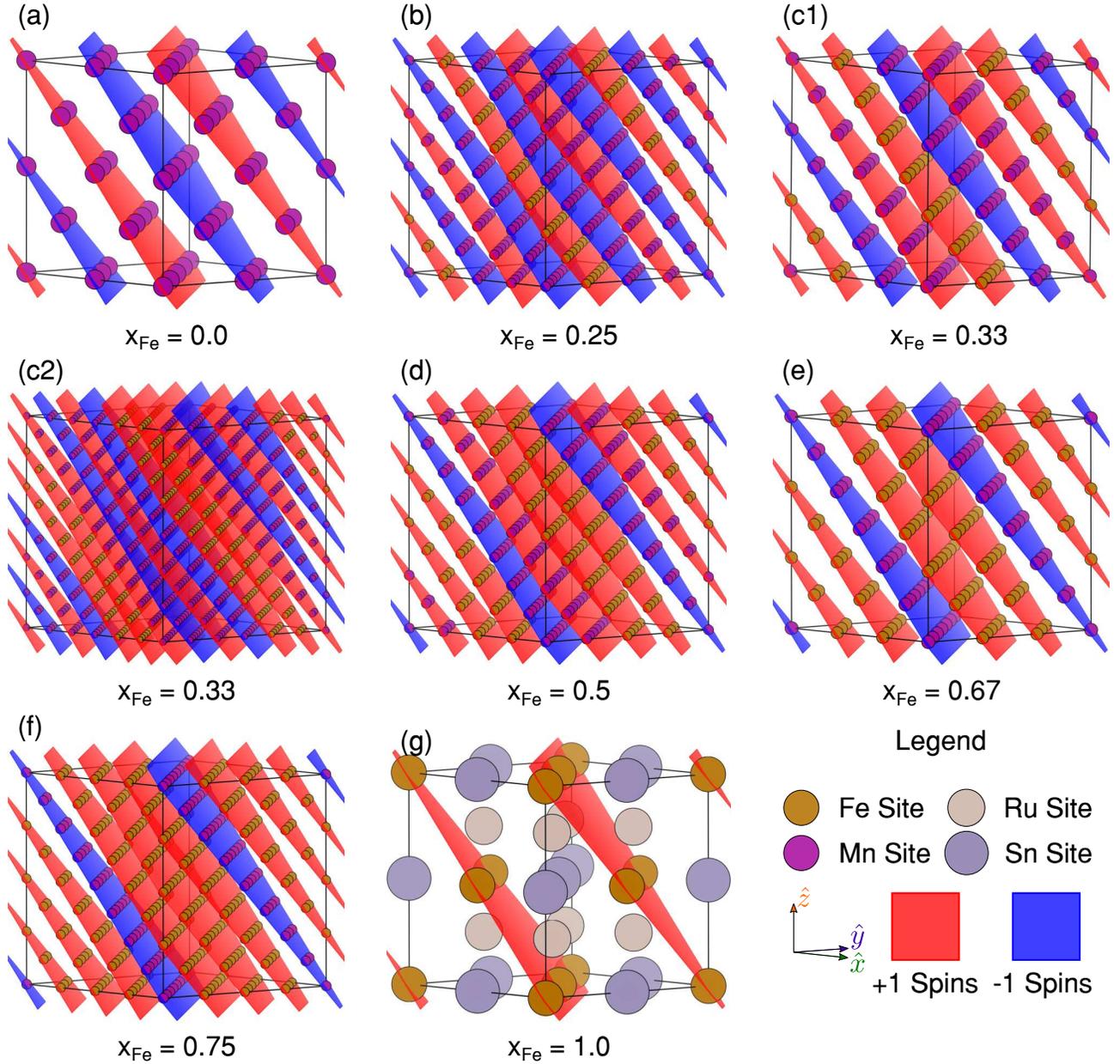}%
\caption{The lowest energy chemical and magnetic configurations for Mn$_{1-x}$Fe$_x$Ru$_2$Sn for six compositions, showing only the (Mn,Fe)/(Spin Up,Spin Down) sites, except for (g), which shows the full unit cell including Ru and Sn. (c1) and (c2) both represent $x_{Fe}$ = 0.33, however, magnetic frustration caused by finite size effects in (c1) results in a significantly larger per-atom energy than the alternate but larger structure (c2). Both chemical and magnetic orderings occur along the [111] direction, creating L1$_1$-like variants across composition space. All visualization were created using the VESTA software package.\label{struc}}%
\end{figure*}

The transition from FM to FrM to AFM is best understood by examining the specific chemical and magnetic configurations of the lowest energy structures as a function of composition.
These structures are shown in Figures~\ref{struc}(a-g), in order of increasing Fe content.
Most striking is the consistency in the ordering: in all cases, variants of the L1$_1$ ordering, consisting of different frequencies of (111) planes that are each chemically and magnetically uniform, were found to have the lowest energy, with Mn-rich regions inducing local AFM behavior.
As seen in Figures~\ref{struc}e and f, Mn aligns AFM to neighboring Fe atoms, even when Mn is dilute.
These low energy AFM arrangements are facilitated via long-range exchange interactions mediated by the Sn sites.
The reverse is not true: in Figure~\ref{struc} (b), the dilute Fe do not induce any local FM ordering in the neighboring Mn planes, resulting in the near-zero net magnetic moments observed at Mn-rich compositions in Figure~\ref{hull} (a).
This suggests that in dilute solid solutions the isolated Fe atoms are forced into the AFM configuration favored by the more populous Mn. The price paid for forcing a Mn-dominated structure to accommodate FM behavior is illustrated by the $>10$ meV/atom energy difference between two structures at $x_{Fe} = 0.33$ having the same chemical ordering but a different magnetic ordering. These structures are illustrated in Figure~\ref{struc} (c1) and (c2). They differ in the size of their magnetic supercell (three and six unit cells, respectively). In the larger supercell, a smaller fraction of the Mn atoms are forced to participate in FM-like behavior, leading to a lower formation energy.


\begin{table}
    \caption{Spin-flip energies in the lowest-energy chemical configurations of Mn$_{1-x}$Fe$_x$Ru$_2$Sn as a function of composition, in meV/($\mu_B$) per primitive cell. The energy differences are calculated as the absolute value of the difference of the energy between the lowest-energy structure at the given $x_{Fe}$, and the pure FM or AFM variant of that structure, normalized per difference in relaxed magnetic moment and per primitive cell.\label{t1}}
    \begin{ruledtabular}
    \begin{tabular}{d c c c c c}
        \multicolumn{1}{c}{$x_{Fe}$} & \multicolumn{1}{c}{$E_{form}$} & \multicolumn{2}{c}{FrM vs. FM}                                                                     & \multicolumn{2}{c}{FrM vs. AFM} \\
                                     & \multicolumn{1}{c}{(meV)}      & \multicolumn{1}{c}{$\Delta \mu_B$ } &  \multicolumn{1}{c}{$\Delta E$} & \multicolumn{1}{c}{$\Delta \mu_B$ } &  \multicolumn{1}{c}{$\Delta E$} \\
                                     &                                & \multicolumn{1}{c}{$(\mu_B)$ } &  \multicolumn{1}{c}{(meV/$\mu_B$) } & \multicolumn{1}{c}{$(\mu_B)$ } &  \multicolumn{1}{c}{(meV/$\mu_B$) } \\ \hline
            0.25  &  12.8  &  3.1  &   9.4  &   0\footnote{For $x_{Fe} = 0.25, 0.33$, the lowest-energy structures are AFM\@.}   &  0$^{\text{a}}$\\
            0.33  &  30.1  &  3.4  &   8.1  &  0$^{\text{a}}$ & 0$^{\text{a}}$ \\
            0.5   &  18.7  &  1.5  &   6.3  & 2.1  & 7.2 \\
            0.67  &  15.8  &  2.0  &   5.5  & 1.7  & 14.7 \\
            0.75  &  12.8  &  1.5  &   4.5  & 2.1 & 20.0 \\
    \end{tabular}
    \end{ruledtabular}
\end{table}

For each of the low energy chemical configurations, the pure-FM and pure-L1$_1$ AFM magnetic configurations are strictly higher in energy than the FrM magnetic configurations shown in Figures~\ref{struc} (b-f).
The energy differences in Table~\ref{t1}, normalized per Bohr magneton, demonstrate that transitions to FM configurations are more costly in Mn-dominated structures, while transitions to AFM are more costly in Fe-dominated structures

\begin{figure*}
    \includegraphics[width=0.95 \textwidth]{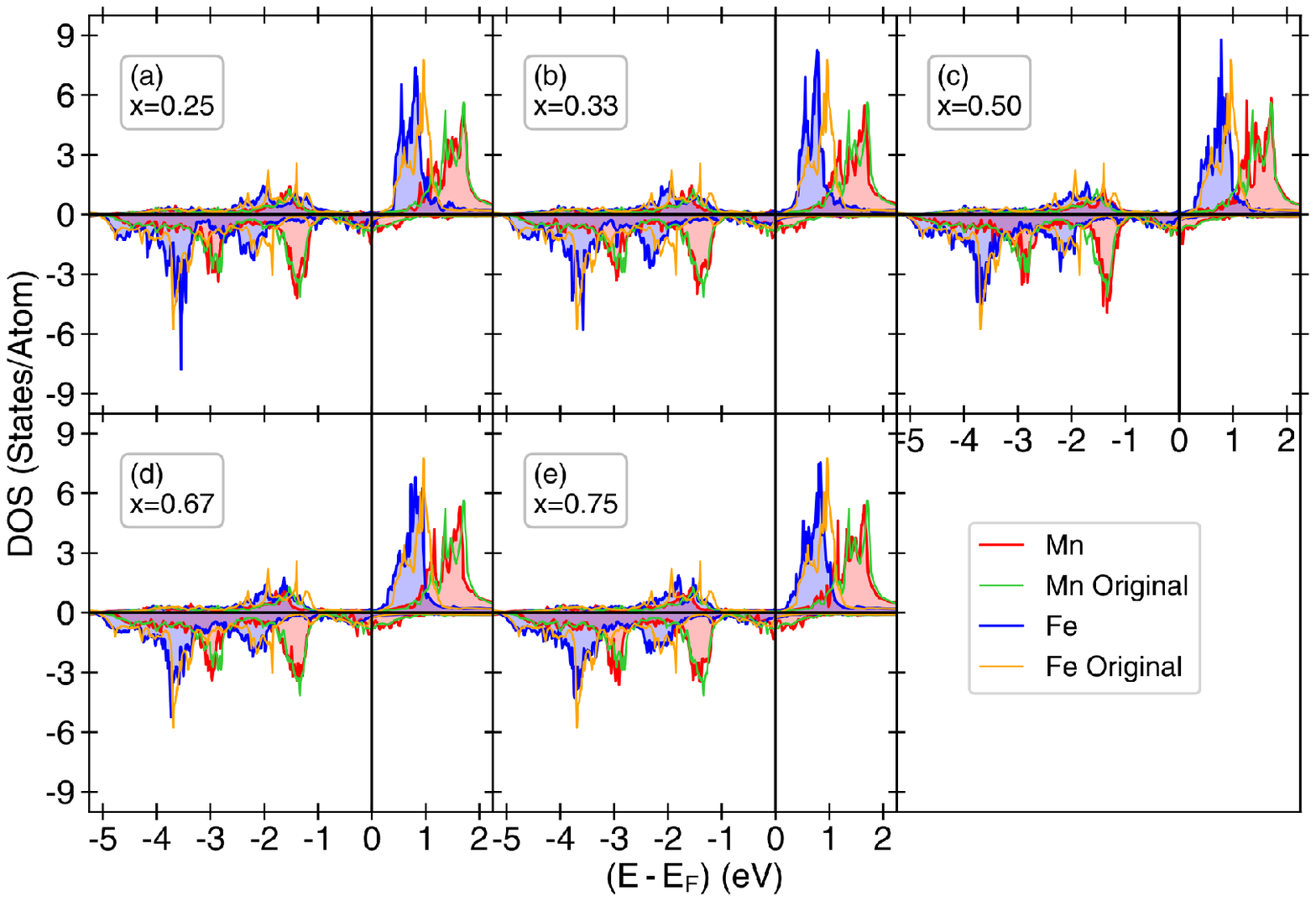}%
    \caption{Site-projected DOS for the Mn and Fe sites in Mn$_{1-x}$Fe$_x$Ru$_2$Sn, compared against site-projected DOS of Mn and Fe sites in the pure Heuslers (Mn,Fe)Ru$_2$Sn. Thin green/orange lines indicate DOS from the pure Heuslers, while thick red/blue lines with shading indicate the Mn or Fe site-projected DOS from the mixed phases.\label{multidos}}%
\end{figure*}

The calculated magnetic moments in the low energy configurations imply that in a well-mixed disordered solid solution, Fe will remain FM and Mn will remain AFM, yielding the mixed magnetic phase proposed by Douglas et al.
We can further confirm this assertion by examining how the electronic structures of the Mn and Fe species in mixed Mn$_{1-x}$Fe$_x$Ru$_2$Sn Heusler alloys compare to the electronic structures of Mn and Fe in pure MnRu$_2$Sn and FeRu$_2$Sn, respectively. Figure~\ref{multidos} shows the site-projected DOS for Mn and Fe in both the pure Heuslers and at each of the intermediate compositions of Mn$_{1-x}$Fe$_x$Ru$_2$Sn.
The DOS of the Mn and Fe sites only change negligibly in the mixed phases, suggesting that the Mn and Fe sites behave similarly in the mixed solid solution and in the pure Heuslers. Thus, we can utilize the analysis of (Mn,Fe)Ru$_2$Sn carried out in Subsection~\ref{pure} on the Heusler alloy Mn$_{1-x}$Fe$_x$Ru$_2$Sn.

Our calculated results, combined with the experimental observations of Douglas et al., lead us to a model of isolated magnetic nano-domains of AFM character embedded in a larger FM matrix for Fe concentrations of $0.5 \leq x < 1$.
We believe the magnetic hardening can be explained by an exchange-hardening effect, where AFM domains centered on individual Mn atoms or small clusters ($<5$ atoms, as no mesoscale ordering is observed experimentally) couple with a single, bulk FM domain carried collectively by the Fe atoms.
Such nano-domains would fall below the observation limit of the experimental techniques used to characterize the chemical and magnetic distributions, and offer the simplest phenomenological explanation of the measured magnetic properties.

\begin{figure}
    \includegraphics[width=0.49 \textwidth]{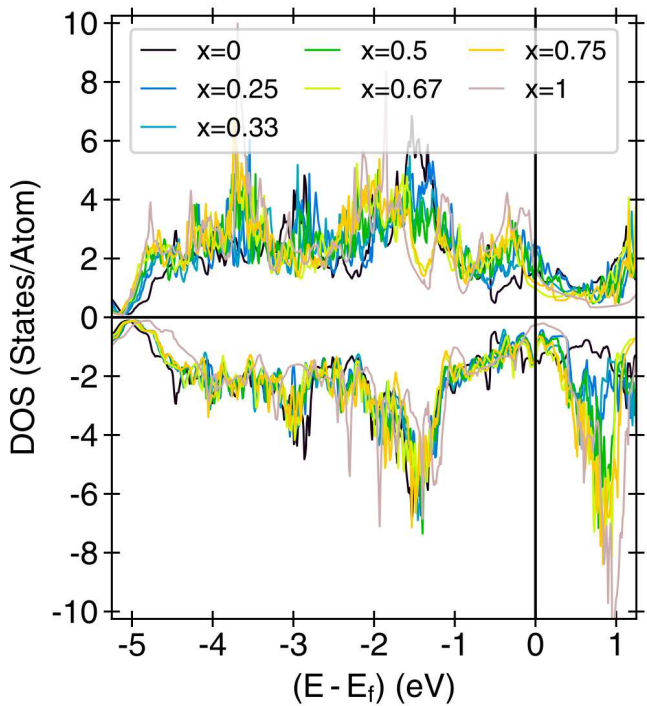}%
    \caption{Total DOS for various compositions of Mn$_{1-x}$Fe$_2$Ru$_2$Sn, showing several ``near-gap'' occupations in the minority spin channel, but no true half-metals.\label{dos}}%
\end{figure}

The Mn and Fe DOS of Figure~\ref{multidos} suggest the possibility of half-metallic behavior.
Half-metallic behavior is (relatively) common among Heuslers, showing up in multiple members of the X(Co,Ni,Mn)$_2$Z families of alloys\cite{Ishida1995,Galanakis2002b,Sasoglu2005,Galanakis2006}. For spintronics, half-metallicity is considered a promising route to achieve the necessary spin-polarized currents. Antiferromagnetic (or rather, fully-compensated ferrimagnetic) half-metals are of particular interest for their net-zero (macroscopic) magnetization, increasing the range of their potential applications\cite{Galanakis2007}. The total DOS for each of the lowest-energy intermediate composition structures is given in Figure~\ref{dos}; in no cases is a true gap achieved in either spin channel. Previous \textit{ab-initio} calculations of the electronic properties of MnRu$_2$Sn indicate a gap in the minority-spin DOS at and around the Fermi level; however, these calculations were performed on the ferromagnetic configuration\cite{Ishida1995a}. As is clear in Figure~\ref{dos} (a), there is no such gap in the AFM ground state\footnote{While the choice of functional (\textit{e.g.}, GGA-PBE vs. LDA) is known to influence the \textit{size} of the gap, the choice of functional should not impact whether a gap is observed at all.\cite{Rai2016}}. 


\section{Conclusion}
We have performed a systematic first-principles investigation of both the chemical and magnetic degrees of freedom in the Mn$_{1-x}$Fe$_{x}$Ru$_{2}$Sn Heusler alloy to determine the origins of exchange hardening in a single phase that is chemically uniform. Careful analysis of the electronic structure of the MnRu$_2$Sn and FeRu$_2$Sn end members has shown that magnetic ordering phenomena on the Mn/Fe sub-lattice is determined by a competition between Sn-mediated superexchange, and conduction-electron-mediated RKKY exchange, with Ru serving only to set the lattice constant. Our results demonstrate that the transition from L1$_1$ antiferromagnetic ordering in MnRu$_2$Sn to ferromagnetic ordering in FeRu$_2$Sn can be explained by a shifting and narrowing of \textit{d}-states near the Fermi level. In MnRu$_2$Sn, the Fermi level bisects a \textit{d}-peak, leaving more states immediately above the Fermi level than below as to favor superexchange, while in FeRu$_2$Sn, the Fermi level falls immediately after a \textit{d}-peak, favoring RKKY-type exchange.


The magnetic ordering phenomena of the pure Heuslers (Mn,Fe)Ru$_2$Sn are not disrupted when Mn and Fe are mixed. First-principles calculations predict that the Mn$_{1-x}$Fe$_{x}$Ru$_{2}$Sn alloy should phase separate in the thermodynamic limit. Low energy Mn/Fe orderings at intermediate concentrations consist of alternating (111) layers that are chemically and magnetically uniform. The calculated total magnetic moments of the intermediate phases follow the trend of moments measured in quenched solid solutions. An analysis of the density of states of low energy structures confirms that the conclusions for pure Heuslers (Mn,Fe)Ru$_2$Sn remain valid at intermediate compositions. While several configurations present a near-gap in the minority spin channel, no true half-metals were discovered in Mn$_{1- 2  x}$Fe$_x$Ru$_2$Sn.

\begin{acknowledgments}
We thank Emily Levin, John Goiri, and Anirudh Natarajan for many helpful discussions and assistance with performing calculations. We also thank Dr.\ Nathan (Fuzzy) Rogers, Dr.\ Paul Weakliem, and Dr.\ Burak Himmetoglu for help with computational facilities.
The research reported here was supported by the Materials Research Science and Engineering Center at UCSB (MRSEC NSF DMR 1720256) through IRG-1.
Simulations were performed using resources from the Center for Scientific Computing in the CNSI and MRL, funded by National Science Foundation MRSEC (DMR-1121053), National Science Foundation CNS-0960316, and Hewlett Packard.
\end{acknowledgments}

\bibliography{library.bib}

\begin{thebibliography}{60}%
\makeatletter
\providecommand \@ifxundefined [1]{%
 \@ifx{#1\undefined}
}%
\providecommand \@ifnum [1]{%
 \ifnum #1\expandafter \@firstoftwo
 \else \expandafter \@secondoftwo
 \fi
}%
\providecommand \@ifx [1]{%
 \ifx #1\expandafter \@firstoftwo
 \else \expandafter \@secondoftwo
 \fi
}%
\providecommand \natexlab [1]{#1}%
\providecommand \enquote  [1]{``#1''}%
\providecommand \bibnamefont  [1]{#1}%
\providecommand \bibfnamefont [1]{#1}%
\providecommand \citenamefont [1]{#1}%
\providecommand \href@noop [0]{\@secondoftwo}%
\providecommand \href [0]{\begingroup \@sanitize@url \@href}%
\providecommand \@href[1]{\@@startlink{#1}\@@href}%
\providecommand \@@href[1]{\endgroup#1\@@endlink}%
\providecommand \@sanitize@url [0]{\catcode `\\12\catcode `\$12\catcode
  `\&12\catcode `\#12\catcode `\^12\catcode `\_12\catcode `\%12\relax}%
\providecommand \@@startlink[1]{}%
\providecommand \@@endlink[0]{}%
\providecommand \url  [0]{\begingroup\@sanitize@url \@url }%
\providecommand \@url [1]{\endgroup\@href {#1}{\urlprefix }}%
\providecommand \urlprefix  [0]{URL }%
\providecommand \Eprint [0]{\href }%
\providecommand \doibase [0]{http://dx.doi.org/}%
\providecommand \selectlanguage [0]{\@gobble}%
\providecommand \bibinfo  [0]{\@secondoftwo}%
\providecommand \bibfield  [0]{\@secondoftwo}%
\providecommand \translation [1]{[#1]}%
\providecommand \BibitemOpen [0]{}%
\providecommand \bibitemStop [0]{}%
\providecommand \bibitemNoStop [0]{.\EOS\space}%
\providecommand \EOS [0]{\spacefactor3000\relax}%
\providecommand \BibitemShut  [1]{\csname bibitem#1\endcsname}%
\let\auto@bib@innerbib\@empty
\bibitem [{\citenamefont {Graf}\ \emph
  {et~al.}(2011{\natexlab{a}})\citenamefont {Graf}, \citenamefont {Felser},\
  and\ \citenamefont {Parkin}}]{Graf2011}%
  \BibitemOpen
  \bibfield  {author} {\bibinfo {author} {\bibfnamefont {T.}~\bibnamefont
  {Graf}}, \bibinfo {author} {\bibfnamefont {C.}~\bibnamefont {Felser}}, \ and\
  \bibinfo {author} {\bibfnamefont {S.~S.}\ \bibnamefont {Parkin}},\ }\href
  {\doibase 10.1016/j.progsolidstchem.2011.02.001} {\bibfield  {journal}
  {\bibinfo  {journal} {Prog. Solid State Chem.}\ }\textbf {\bibinfo {volume}
  {39}},\ \bibinfo {pages} {1} (\bibinfo {year}
  {2011}{\natexlab{a}})}\BibitemShut {NoStop}%
\bibitem [{\citenamefont {Graf}\ \emph
  {et~al.}(2011{\natexlab{b}})\citenamefont {Graf}, \citenamefont {Parkin},\
  and\ \citenamefont {Felser}}]{Graf2011a}%
  \BibitemOpen
  \bibfield  {author} {\bibinfo {author} {\bibfnamefont {T.}~\bibnamefont
  {Graf}}, \bibinfo {author} {\bibfnamefont {S.~S.~P.}\ \bibnamefont {Parkin}},
  \ and\ \bibinfo {author} {\bibfnamefont {C.}~\bibnamefont {Felser}},\ }\href
  {\doibase 10.1109/TMAG.2010.2096229} {\bibfield  {journal} {\bibinfo
  {journal} {IEEE Trans. Magn.}\ }\textbf {\bibinfo {volume} {47}},\ \bibinfo
  {pages} {367} (\bibinfo {year} {2011}{\natexlab{b}})}\BibitemShut {NoStop}%
\bibitem [{\citenamefont {Planes}\ \emph {et~al.}(2009)\citenamefont {Planes},
  \citenamefont {Ma{\~{n}}osa},\ and\ \citenamefont {Acet}}]{Planes2009}%
  \BibitemOpen
  \bibfield  {author} {\bibinfo {author} {\bibfnamefont {A.}~\bibnamefont
  {Planes}}, \bibinfo {author} {\bibfnamefont {L.}~\bibnamefont
  {Ma{\~{n}}osa}}, \ and\ \bibinfo {author} {\bibfnamefont {M.}~\bibnamefont
  {Acet}},\ }\href {\doibase 10.1088/0953-8984/21/23/233201} {\bibfield
  {journal} {\bibinfo  {journal} {J. Phys. Condens. Matter}\ }\textbf {\bibinfo
  {volume} {21}},\ \bibinfo {pages} {1} (\bibinfo {year} {2009})}\BibitemShut
  {NoStop}%
\bibitem [{\citenamefont {Yin}\ \emph {et~al.}(2016)\citenamefont {Yin},
  \citenamefont {Hasier},\ and\ \citenamefont {Nash}}]{Yin2015}%
  \BibitemOpen
  \bibfield  {author} {\bibinfo {author} {\bibfnamefont {M.}~\bibnamefont
  {Yin}}, \bibinfo {author} {\bibfnamefont {J.}~\bibnamefont {Hasier}}, \ and\
  \bibinfo {author} {\bibfnamefont {P.}~\bibnamefont {Nash}},\ }\href {\doibase
  10.1007/s10853-015-9389-y} {\bibfield  {journal} {\bibinfo  {journal} {J.
  Mater. Sci.}\ }\textbf {\bibinfo {volume} {51}},\ \bibinfo {pages} {50}
  (\bibinfo {year} {2016})}\BibitemShut {NoStop}%
\bibitem [{\citenamefont {Stearns}(1980)}]{Stearns1980}%
  \BibitemOpen
  \bibfield  {author} {\bibinfo {author} {\bibfnamefont {M.~B.}\ \bibnamefont
  {Stearns}},\ }\href {\doibase 10.1016/0304-8853(80)91060-4} {\bibfield
  {journal} {\bibinfo  {journal} {J. Magn. Magn. Mater.}\ }\textbf {\bibinfo
  {volume} {15-18}},\ \bibinfo {pages} {301} (\bibinfo {year}
  {1980})}\BibitemShut {NoStop}%
\bibitem [{\citenamefont {K{\"{u}}bler}\ \emph {et~al.}(1983)\citenamefont
  {K{\"{u}}bler}, \citenamefont {Williams},\ and\ \citenamefont
  {Sommers}}]{Kubler1983}%
  \BibitemOpen
  \bibfield  {author} {\bibinfo {author} {\bibfnamefont {J.}~\bibnamefont
  {K{\"{u}}bler}}, \bibinfo {author} {\bibfnamefont {A.~R.}\ \bibnamefont
  {Williams}}, \ and\ \bibinfo {author} {\bibfnamefont {C.~B.}\ \bibnamefont
  {Sommers}},\ }\href {\doibase 10.1103/PhysRevB.28.1745} {\bibfield  {journal}
  {\bibinfo  {journal} {Phys. Rev. B}\ }\textbf {\bibinfo {volume} {28}},\
  \bibinfo {pages} {1745} (\bibinfo {year} {1983})},\ \Eprint
  {http://arxiv.org/abs/0411737v2} {arXiv:0411737v2 [cond-mat]} \BibitemShut
  {NoStop}%
\bibitem [{\citenamefont {Kurtulus}\ \emph {et~al.}(2005)\citenamefont
  {Kurtulus}, \citenamefont {Dronskowski}, \citenamefont {Samolyuk},\ and\
  \citenamefont {Antropov}}]{Kurtulus2005}%
  \BibitemOpen
  \bibfield  {author} {\bibinfo {author} {\bibfnamefont {Y.}~\bibnamefont
  {Kurtulus}}, \bibinfo {author} {\bibfnamefont {R.}~\bibnamefont
  {Dronskowski}}, \bibinfo {author} {\bibfnamefont {G.~D.}\ \bibnamefont
  {Samolyuk}}, \ and\ \bibinfo {author} {\bibfnamefont {V.~P.}\ \bibnamefont
  {Antropov}},\ }\href {\doibase 10.1103/PhysRevB.71.014425} {\bibfield
  {journal} {\bibinfo  {journal} {Phys. Rev. B}\ }\textbf {\bibinfo {volume}
  {71}},\ \bibinfo {pages} {014425} (\bibinfo {year} {2005})},\ \Eprint
  {http://arxiv.org/abs/0406588} {arXiv:0406588 [cond-mat]} \BibitemShut
  {NoStop}%
\bibitem [{\citenamefont {Kurtulus}\ \emph {et~al.}(2006)\citenamefont
  {Kurtulus}, \citenamefont {Gille{\ss}en},\ and\ \citenamefont
  {Dronskowski}}]{Kurtulus2006}%
  \BibitemOpen
  \bibfield  {author} {\bibinfo {author} {\bibfnamefont {Y.}~\bibnamefont
  {Kurtulus}}, \bibinfo {author} {\bibfnamefont {M.}~\bibnamefont
  {Gille{\ss}en}}, \ and\ \bibinfo {author} {\bibfnamefont {R.}~\bibnamefont
  {Dronskowski}},\ }\href {\doibase 10.1002/jcc.20308} {\bibfield  {journal}
  {\bibinfo  {journal} {J. Comput. Chem.}\ }\textbf {\bibinfo {volume} {27}},\
  \bibinfo {pages} {90} (\bibinfo {year} {2006})}\BibitemShut {NoStop}%
\bibitem [{\citenamefont {Galanakis}\ \emph
  {et~al.}(2002{\natexlab{a}})\citenamefont {Galanakis}, \citenamefont
  {Dederichs},\ and\ \citenamefont {Papanikolaou}}]{Galanakis2002b}%
  \BibitemOpen
  \bibfield  {author} {\bibinfo {author} {\bibfnamefont {I.}~\bibnamefont
  {Galanakis}}, \bibinfo {author} {\bibfnamefont {P.~H.}\ \bibnamefont
  {Dederichs}}, \ and\ \bibinfo {author} {\bibfnamefont {N.}~\bibnamefont
  {Papanikolaou}},\ }\href {\doibase 10.1103/PhysRevB.66.174429} {\bibfield
  {journal} {\bibinfo  {journal} {Phys. Rev. B}\ }\textbf {\bibinfo {volume}
  {66}},\ \bibinfo {pages} {174429} (\bibinfo {year}
  {2002}{\natexlab{a}})}\BibitemShut {NoStop}%
\bibitem [{\citenamefont {Galanakis}\ \emph {et~al.}(2006)\citenamefont
  {Galanakis}, \citenamefont {Mavropoulos},\ and\ \citenamefont
  {Dederichs}}]{Galanakis2006}%
  \BibitemOpen
  \bibfield  {author} {\bibinfo {author} {\bibfnamefont {I.}~\bibnamefont
  {Galanakis}}, \bibinfo {author} {\bibfnamefont {P.}~\bibnamefont
  {Mavropoulos}}, \ and\ \bibinfo {author} {\bibfnamefont {P.~H.}\ \bibnamefont
  {Dederichs}},\ }\href {\doibase 10.1088/0022-3727/39/5/S01} {\bibfield
  {journal} {\bibinfo  {journal} {J. Phys. D. Appl. Phys.}\ }\textbf {\bibinfo
  {volume} {39}},\ \bibinfo {pages} {765} (\bibinfo {year} {2006})},\ \Eprint
  {http://arxiv.org/abs/0510276} {arXiv:0510276 [cond-mat]} \BibitemShut
  {NoStop}%
\bibitem [{\citenamefont {Ishida}\ \emph
  {et~al.}(1995{\natexlab{a}})\citenamefont {Ishida}, \citenamefont
  {Kashiwagi}, \citenamefont {Fujii},\ and\ \citenamefont
  {Asano}}]{Ishida1995a}%
  \BibitemOpen
  \bibfield  {author} {\bibinfo {author} {\bibfnamefont {S.}~\bibnamefont
  {Ishida}}, \bibinfo {author} {\bibfnamefont {S.}~\bibnamefont {Kashiwagi}},
  \bibinfo {author} {\bibfnamefont {S.}~\bibnamefont {Fujii}}, \ and\ \bibinfo
  {author} {\bibfnamefont {S.}~\bibnamefont {Asano}},\ }\href {\doibase
  10.1016/0921-4526(94)00920-Q} {\bibfield  {journal} {\bibinfo  {journal}
  {Phys. B Phys. Condens. Matter}\ }\textbf {\bibinfo {volume} {210}},\
  \bibinfo {pages} {140} (\bibinfo {year} {1995}{\natexlab{a}})}\BibitemShut
  {NoStop}%
\bibitem [{\citenamefont {\c{S}a\c{s}{\i}o\u{g}lu}\ \emph {et~al.}(2005)\citenamefont
  {\c{S}a\c{s}{\i}o\u{g}lu}, \citenamefont {Sandratskii}, \citenamefont {Bruno},\ and\
  \citenamefont {Galanakis}}]{Sasoglu2005}%
  \BibitemOpen
  \bibfield  {author} {\bibinfo {author} {\bibfnamefont {E.}~\bibnamefont
  {\c{S}a\c{s}{\i}o\u{g}lu}}, \bibinfo {author} {\bibfnamefont {L.~M.}\ \bibnamefont
  {Sandratskii}}, \bibinfo {author} {\bibfnamefont {P.}~\bibnamefont {Bruno}},
  \ and\ \bibinfo {author} {\bibfnamefont {I.}~\bibnamefont {Galanakis}},\
  }\href {\doibase 10.1103/PhysRevB.72.184415} {\bibfield  {journal} {\bibinfo
  {journal} {Phys. Rev. B}\ }\textbf {\bibinfo {volume} {72}},\ \bibinfo
  {pages} {184415} (\bibinfo {year} {2005})}\BibitemShut {NoStop}%
\bibitem [{\citenamefont {Mizusaki}\ \emph {et~al.}(2011)\citenamefont
  {Mizusaki}, \citenamefont {Douzono}, \citenamefont {Ohnishi}, \citenamefont
  {Ozawa}, \citenamefont {Samata}, \citenamefont {Noro},\ and\ \citenamefont
  {Nagata}}]{Mizusaki2011}%
  \BibitemOpen
  \bibfield  {author} {\bibinfo {author} {\bibfnamefont {S.}~\bibnamefont
  {Mizusaki}}, \bibinfo {author} {\bibfnamefont {A.}~\bibnamefont {Douzono}},
  \bibinfo {author} {\bibfnamefont {T.}~\bibnamefont {Ohnishi}}, \bibinfo
  {author} {\bibfnamefont {T.~C.}\ \bibnamefont {Ozawa}}, \bibinfo {author}
  {\bibfnamefont {H.}~\bibnamefont {Samata}}, \bibinfo {author} {\bibfnamefont
  {Y.}~\bibnamefont {Noro}}, \ and\ \bibinfo {author} {\bibfnamefont
  {Y.}~\bibnamefont {Nagata}},\ }\href {\doibase 10.1016/j.jallcom.2011.09.017}
  {\bibfield  {journal} {\bibinfo  {journal} {J. Alloys Compd.}\ }\textbf
  {\bibinfo {volume} {510}},\ \bibinfo {pages} {141} (\bibinfo {year}
  {2011})}\BibitemShut {NoStop}%
\bibitem [{\citenamefont {Douglas}\ \emph {et~al.}(2016)\citenamefont
  {Douglas}, \citenamefont {Levin}, \citenamefont {Pollock}, \citenamefont
  {Castillo}, \citenamefont {Adler}, \citenamefont {Felser}, \citenamefont
  {Kr{\"{a}}mer}, \citenamefont {Page},\ and\ \citenamefont
  {Seshadri}}]{Douglas2016}%
  \BibitemOpen
  \bibfield  {author} {\bibinfo {author} {\bibfnamefont {J.~E.}\ \bibnamefont
  {Douglas}}, \bibinfo {author} {\bibfnamefont {E.~E.}\ \bibnamefont {Levin}},
  \bibinfo {author} {\bibfnamefont {T.~M.}\ \bibnamefont {Pollock}}, \bibinfo
  {author} {\bibfnamefont {J.~C.}\ \bibnamefont {Castillo}}, \bibinfo {author}
  {\bibfnamefont {P.}~\bibnamefont {Adler}}, \bibinfo {author} {\bibfnamefont
  {C.}~\bibnamefont {Felser}}, \bibinfo {author} {\bibfnamefont
  {S.}~\bibnamefont {Kr{\"{a}}mer}}, \bibinfo {author} {\bibfnamefont {K.~L.}\
  \bibnamefont {Page}}, \ and\ \bibinfo {author} {\bibfnamefont
  {R.}~\bibnamefont {Seshadri}},\ }\href {\doibase 10.1103/PhysRevB.94.094412}
  {\bibfield  {journal} {\bibinfo  {journal} {Phys. Rev. B}\ }\textbf {\bibinfo
  {volume} {94}},\ \bibinfo {pages} {094412} (\bibinfo {year}
  {2016})}\BibitemShut {NoStop}%
\bibitem [{\citenamefont {Pebley}\ \emph {et~al.}(2016)\citenamefont {Pebley},
  \citenamefont {Fuks}, \citenamefont {Pollock},\ and\ \citenamefont
  {Gordon}}]{Pebley2016}%
  \BibitemOpen
  \bibfield  {author} {\bibinfo {author} {\bibfnamefont {A.~C.}\ \bibnamefont
  {Pebley}}, \bibinfo {author} {\bibfnamefont {P.~E.}\ \bibnamefont {Fuks}},
  \bibinfo {author} {\bibfnamefont {T.~M.}\ \bibnamefont {Pollock}}, \ and\
  \bibinfo {author} {\bibfnamefont {M.~J.}\ \bibnamefont {Gordon}},\ }\href
  {\doibase 10.1016/j.jmmm.2016.06.009} {\bibfield  {journal} {\bibinfo
  {journal} {J. Magn. Magn. Mater.}\ }\textbf {\bibinfo {volume} {419}},\
  \bibinfo {pages} {29} (\bibinfo {year} {2016})}\BibitemShut {NoStop}%
\bibitem [{\citenamefont {Pebley}\ \emph {et~al.}(2017)\citenamefont {Pebley},
  \citenamefont {Decolvenaere}, \citenamefont {Pollock},\ and\ \citenamefont
  {Gordon}}]{Pebley2017}%
  \BibitemOpen
  \bibfield  {author} {\bibinfo {author} {\bibfnamefont {A.}~\bibnamefont
  {Pebley}}, \bibinfo {author} {\bibfnamefont {E.}~\bibnamefont
  {Decolvenaere}}, \bibinfo {author} {\bibfnamefont {T.~M.}\ \bibnamefont
  {Pollock}}, \ and\ \bibinfo {author} {\bibfnamefont {M.}~\bibnamefont
  {Gordon}},\ }\href {\doibase 10.1039/C7NR04302C} {\bibfield  {journal}
  {\bibinfo  {journal} {Nanoscale}\ } (\bibinfo {year} {2017}),\
  10.1039/C7NR04302C}\BibitemShut {NoStop}%
\bibitem [{\citenamefont {{Hohenberg, P.; Kohn}}(1964)}]{Hohenberg1964}%
  \BibitemOpen
  \bibfield  {author} {\bibinfo {author} {\bibfnamefont {W.}~\bibnamefont
  {{Hohenberg, P.; Kohn}}},\ }\href {\doibase 10.1103/PhysRev.136.B864}
  {\bibfield  {journal} {\bibinfo  {journal} {Phys. Rev.}\ }\textbf {\bibinfo
  {volume} {136}},\ \bibinfo {pages} {B864} (\bibinfo {year} {1964})},\ \Eprint
  {http://arxiv.org/abs/1108.5632} {arXiv:1108.5632} \BibitemShut {NoStop}%
\bibitem [{\citenamefont {Kohn}\ and\ \citenamefont {Sham}(1965)}]{Kohn1965}%
  \BibitemOpen
  \bibfield  {author} {\bibinfo {author} {\bibfnamefont {W.}~\bibnamefont
  {Kohn}}\ and\ \bibinfo {author} {\bibfnamefont {L.~J.}\ \bibnamefont
  {Sham}},\ }\href {\doibase 10.1103/PhysRev.140.A1133} {\bibfield  {journal}
  {\bibinfo  {journal} {Phys. Rev.}\ }\textbf {\bibinfo {volume} {140}},\
  \bibinfo {pages} {A1133} (\bibinfo {year} {1965})},\ \Eprint
  {http://arxiv.org/abs/PhysRev.140.A1133} {arXiv:PhysRev.140.A1133 [10.1103]}
  \BibitemShut {NoStop}%
\bibitem [{\citenamefont {Heusler}(1903)}]{Heusler1903}%
  \BibitemOpen
  \bibfield  {author} {\bibinfo {author} {\bibfnamefont {F.}~\bibnamefont
  {Heusler}},\ }\href@noop {} {\bibfield  {journal} {\bibinfo  {journal}
  {Verhandlungen der Dtsch. Phys. Gesellschaft}\ }\textbf {\bibinfo {volume}
  {5}},\ \bibinfo {pages} {219} (\bibinfo {year} {1903})}\BibitemShut {NoStop}%
\bibitem [{Note1()}]{Note1}%
  \BibitemOpen
  \bibinfo {note} {Antiferromagnetism would not be proposed as a phenomena for
  another 30 years\cite {Neel1936}}\BibitemShut {NoStop}%
\bibitem [{\citenamefont {Galanakis}\ \emph
  {et~al.}(2002{\natexlab{b}})\citenamefont {Galanakis}, \citenamefont
  {Dederichs},\ and\ \citenamefont {Papanikolaou}}]{Galanakis2002}%
  \BibitemOpen
  \bibfield  {author} {\bibinfo {author} {\bibfnamefont {I.}~\bibnamefont
  {Galanakis}}, \bibinfo {author} {\bibfnamefont {P.~H.}\ \bibnamefont
  {Dederichs}}, \ and\ \bibinfo {author} {\bibfnamefont {N.}~\bibnamefont
  {Papanikolaou}},\ }\href {\doibase 10.1103/PhysRevB.66.134428} {\bibfield
  {journal} {\bibinfo  {journal} {Phys. Rev. B}\ }\textbf {\bibinfo {volume}
  {66}},\ \bibinfo {pages} {134428} (\bibinfo {year}
  {2002}{\natexlab{b}})}\BibitemShut {NoStop}%
\bibitem [{\citenamefont {Galanakis}\ and\ \citenamefont
  {Dederichs}(2005)}]{Galanakis2005a}%
  \BibitemOpen
  \bibinfo {editor} {\bibfnamefont {I.}~\bibnamefont {Galanakis}}\ and\
  \bibinfo {editor} {\bibfnamefont {P.}~\bibnamefont {Dederichs}},\ eds.,\
  \href {\doibase 10.1007/b137760} {\emph {\bibinfo {title} {{Half-metallic
  Alloys}}}},\ \bibinfo {series} {Lecture Notes in Physics}, Vol.\ \bibinfo
  {volume} {676}\ (\bibinfo  {publisher} {Springer Berlin Heidelberg},\
  \bibinfo {address} {Berlin, Heidelberg},\ \bibinfo {year} {2005})\BibitemShut
  {NoStop}%
\bibitem [{\citenamefont {Williams}\ \emph {et~al.}(1983)\citenamefont
  {Williams}, \citenamefont {Moruzzi}, \citenamefont {Gelatt},\ and\
  \citenamefont {K{\"{u}}bler}}]{Williams1983}%
  \BibitemOpen
  \bibfield  {author} {\bibinfo {author} {\bibfnamefont {A.~R.}\ \bibnamefont
  {Williams}}, \bibinfo {author} {\bibfnamefont {V.~L.}\ \bibnamefont
  {Moruzzi}}, \bibinfo {author} {\bibfnamefont {C.~D.}\ \bibnamefont {Gelatt}},
  \ and\ \bibinfo {author} {\bibfnamefont {J.}~\bibnamefont {K{\"{u}}bler}},\
  }\href {\doibase 10.1016/0304-8853(83)90166-X} {\bibfield  {journal}
  {\bibinfo  {journal} {J. Magn. Magn. Mater.}\ }\textbf {\bibinfo {volume}
  {31-34}},\ \bibinfo {pages} {88} (\bibinfo {year} {1983})}\BibitemShut
  {NoStop}%
\bibitem [{\citenamefont {\c{S}a\c{s}{\i}o\u{g}lu}\ \emph {et~al.}(2008)\citenamefont
  {\c{S}a\c{s}{\i}o\u{g}lu}, \citenamefont {Sandratskii},\ and\ \citenamefont
  {Bruno}}]{Sasoglu2008}%
  \BibitemOpen
  \bibfield  {author} {\bibinfo {author} {\bibfnamefont {E.}~\bibnamefont
  {\c{S}a\c{s}{\i}o\u{g}lu}}, \bibinfo {author} {\bibfnamefont {L.~M.}\ \bibnamefont
  {Sandratskii}}, \ and\ \bibinfo {author} {\bibfnamefont {P.}~\bibnamefont
  {Bruno}},\ }\href {\doibase 10.1103/PhysRevB.77.064417} {\bibfield  {journal}
  {\bibinfo  {journal} {Phys. Rev. B}\ }\textbf {\bibinfo {volume} {77}},\
  \bibinfo {pages} {064417} (\bibinfo {year} {2008})},\ \Eprint
  {http://arxiv.org/abs/arXiv:0712.0158v1} {arXiv:arXiv:0712.0158v1}
  \BibitemShut {NoStop}%
\bibitem [{\citenamefont {\c{S}a\c{s}{\i}o\u{g}lu}\ \emph {et~al.}(2004)\citenamefont
  {\c{S}a\c{s}{\i}o\u{g}lu}, \citenamefont {Sandratskii},\ and\ \citenamefont
  {Bruno}}]{Sasoglu2004}%
  \BibitemOpen
  \bibfield  {author} {\bibinfo {author} {\bibfnamefont {E.}~\bibnamefont
  {\c{S}a\c{s}{\i}o\u{g}lu}}, \bibinfo {author} {\bibfnamefont {L.~M.}\ \bibnamefont
  {Sandratskii}}, \ and\ \bibinfo {author} {\bibfnamefont {P.}~\bibnamefont
  {Bruno}},\ }\href {\doibase 10.1103/PhysRevB.70.024427} {\bibfield  {journal}
  {\bibinfo  {journal} {Phys. Rev. B}\ }\textbf {\bibinfo {volume} {70}},\
  \bibinfo {pages} {024427} (\bibinfo {year} {2004})},\ \Eprint
  {http://arxiv.org/abs/0404162} {arXiv:0404162 [cond-mat]} \BibitemShut
  {NoStop}%
\bibitem [{\citenamefont {\c{S}a\c{s}{\i}o\u{g}lu}\ \emph {et~al.}(2005)\citenamefont
  {\c{S}a\c{s}{\i}o\u{g}lu}, \citenamefont {Sandratskii},\ and\ \citenamefont
  {Bruno}}]{Sasoglu2005a}%
  \BibitemOpen
  \bibfield  {author} {\bibinfo {author} {\bibfnamefont {E.}~\bibnamefont
  {\c{S}a\c{s}{\i}o\u{g}lu}}, \bibinfo {author} {\bibfnamefont {L.~M.}\ \bibnamefont
  {Sandratskii}}, \ and\ \bibinfo {author} {\bibfnamefont {P.}~\bibnamefont
  {Bruno}},\ }\href {\doibase 10.1088/0953-8984/17/6/017} {\bibfield  {journal}
  {\bibinfo  {journal} {J. Phys. Condens. Matter}\ }\textbf {\bibinfo {volume}
  {17}},\ \bibinfo {pages} {995} (\bibinfo {year} {2005})},\ \Eprint
  {http://arxiv.org/abs/0504679} {arXiv:0504679 [cond-mat]} \BibitemShut
  {NoStop}%
\bibitem [{Note2()}]{Note2}%
  \BibitemOpen
  \bibinfo {note} {At roughly 30 electrons, Galanakis et al. note that this
  rule begins to break down: the exchange-splitting required to push electrons
  into the highest-energy anti-bonding orbitals is infeasibly
  large}\BibitemShut {NoStop}%
\bibitem [{\citenamefont {Shi}\ \emph {et~al.}(1994)\citenamefont {Shi},
  \citenamefont {Levy},\ and\ \citenamefont {Fry}}]{Shi1994}%
  \BibitemOpen
  \bibfield  {author} {\bibinfo {author} {\bibfnamefont {Z.~P.}\ \bibnamefont
  {Shi}}, \bibinfo {author} {\bibfnamefont {P.~M.}\ \bibnamefont {Levy}}, \
  and\ \bibinfo {author} {\bibfnamefont {J.~L.}\ \bibnamefont {Fry}},\ }\href
  {\doibase 10.1103/PhysRevB.49.15159} {\bibfield  {journal} {\bibinfo
  {journal} {Phys. Rev. B}\ }\textbf {\bibinfo {volume} {49}},\ \bibinfo
  {pages} {15159} (\bibinfo {year} {1994})}\BibitemShut {NoStop}%
\bibitem [{\citenamefont {Azumi}\ and\ \citenamefont
  {Goldman}(1954)}]{Azumi1954}%
  \BibitemOpen
  \bibfield  {author} {\bibinfo {author} {\bibfnamefont {K.}~\bibnamefont
  {Azumi}}\ and\ \bibinfo {author} {\bibfnamefont {J.~E.}\ \bibnamefont
  {Goldman}},\ }\href {\doibase 10.1103/PhysRev.93.630} {\bibfield  {journal}
  {\bibinfo  {journal} {Phys. Rev.}\ }\textbf {\bibinfo {volume} {93}},\
  \bibinfo {pages} {630} (\bibinfo {year} {1954})}\BibitemShut {NoStop}%
\bibitem [{\citenamefont {Kresse}\ and\ \citenamefont
  {Hafner}(1994)}]{Kresse1994}%
  \BibitemOpen
  \bibfield  {author} {\bibinfo {author} {\bibfnamefont {G.}~\bibnamefont
  {Kresse}}\ and\ \bibinfo {author} {\bibfnamefont {J.}~\bibnamefont
  {Hafner}},\ }\href {\doibase 10.1103/PhysRevB.49.14251} {\bibfield  {journal}
  {\bibinfo  {journal} {Phys. Rev. B}\ }\textbf {\bibinfo {volume} {49}},\
  \bibinfo {pages} {14251} (\bibinfo {year} {1994})}\BibitemShut {NoStop}%
\bibitem [{\citenamefont {Kresse}(1996)}]{Kresse1996a}%
  \BibitemOpen
  \bibfield  {author} {\bibinfo {author} {\bibfnamefont {G.}~\bibnamefont
  {Kresse}},\ }\href {\doibase 10.1103/PhysRevB.54.11169} {\bibfield  {journal}
  {\bibinfo  {journal} {Phys. Rev. B}\ }\textbf {\bibinfo {volume} {54}},\
  \bibinfo {pages} {11169} (\bibinfo {year} {1996})}\BibitemShut {NoStop}%
\bibitem [{\citenamefont {Kresse}\ and\ \citenamefont
  {Furthm{\"{u}}ller}(1996)}]{Kresse1996}%
  \BibitemOpen
  \bibfield  {author} {\bibinfo {author} {\bibfnamefont {G.}~\bibnamefont
  {Kresse}}\ and\ \bibinfo {author} {\bibfnamefont {J.}~\bibnamefont
  {Furthm{\"{u}}ller}},\ }\href {\doibase 10.1016/0927-0256(96)00008-0}
  {\bibfield  {journal} {\bibinfo  {journal} {Comput. Mater. Sci.}\ }\textbf
  {\bibinfo {volume} {6}},\ \bibinfo {pages} {15} (\bibinfo {year} {1996})},\
  \Eprint {http://arxiv.org/abs/0927-0256(96)00008} {arXiv:0927-0256(96)00008
  [10.1016]} \BibitemShut {NoStop}%
\bibitem [{\citenamefont {Bl{\"{o}}chl}(1994)}]{Blochl1994a}%
  \BibitemOpen
  \bibfield  {author} {\bibinfo {author} {\bibfnamefont {P.~E.}\ \bibnamefont
  {Bl{\"{o}}chl}},\ }\href {\doibase 10.1103/PhysRevB.50.17953} {\bibfield
  {journal} {\bibinfo  {journal} {Phys. Rev. B}\ }\textbf {\bibinfo {volume}
  {50}},\ \bibinfo {pages} {17953} (\bibinfo {year} {1994})}\BibitemShut
  {NoStop}%
\bibitem [{\citenamefont {Kresse}\ and\ \citenamefont
  {Joubert}(1999)}]{Kresse1999}%
  \BibitemOpen
  \bibfield  {author} {\bibinfo {author} {\bibfnamefont {G.}~\bibnamefont
  {Kresse}}\ and\ \bibinfo {author} {\bibfnamefont {D.}~\bibnamefont
  {Joubert}},\ }\href {\doibase 10.1103/PhysRevB.59.1758} {\bibfield  {journal}
  {\bibinfo  {journal} {Phys. Rev. B}\ }\textbf {\bibinfo {volume} {59}},\
  \bibinfo {pages} {1758} (\bibinfo {year} {1999})}\BibitemShut {NoStop}%
\bibitem [{Note3()}]{Note3}%
  \BibitemOpen
  \bibinfo {note} {Dataset v.54, specific PAWs for each element were chosen
  using the guidelines at \protect \url
  {https://cms.mpi.univie.ac.at/vasp/vasp/Recommended_GW_PAW_potentials_vasp_5_2.html}}\BibitemShut
  {NoStop}%
\bibitem [{\citenamefont {Perdew}\ \emph {et~al.}(1996)\citenamefont {Perdew},
  \citenamefont {Burke},\ and\ \citenamefont {Ernzerhof}}]{Perdew1996}%
  \BibitemOpen
  \bibfield  {author} {\bibinfo {author} {\bibfnamefont {J.~P.}\ \bibnamefont
  {Perdew}}, \bibinfo {author} {\bibfnamefont {K.}~\bibnamefont {Burke}}, \
  and\ \bibinfo {author} {\bibfnamefont {M.}~\bibnamefont {Ernzerhof}},\ }\href
  {\doibase 10.1103/PhysRevLett.77.3865} {\bibfield  {journal} {\bibinfo
  {journal} {Phys. Rev. Lett.}\ }\textbf {\bibinfo {volume} {77}},\ \bibinfo
  {pages} {3865} (\bibinfo {year} {1996})}\BibitemShut {NoStop}%
\bibitem [{\citenamefont {Vosko}\ \emph {et~al.}(1980)\citenamefont {Vosko},
  \citenamefont {Wilk},\ and\ \citenamefont {Nusair}}]{Vosko1980}%
  \BibitemOpen
  \bibfield  {author} {\bibinfo {author} {\bibfnamefont {S.~H.}\ \bibnamefont
  {Vosko}}, \bibinfo {author} {\bibfnamefont {L.}~\bibnamefont {Wilk}}, \ and\
  \bibinfo {author} {\bibfnamefont {M.}~\bibnamefont {Nusair}},\ }\href
  {\doibase 10.1139/p80-159} {\bibfield  {journal} {\bibinfo  {journal} {Can.
  J. Phys.}\ }\textbf {\bibinfo {volume} {58}},\ \bibinfo {pages} {1200}
  (\bibinfo {year} {1980})}\BibitemShut {NoStop}%
\bibitem [{\citenamefont {{Monkhors H.}}\ and\ \citenamefont
  {Pack}(1976)}]{Monkhorst1976}%
  \BibitemOpen
  \bibfield  {author} {\bibinfo {author} {\bibnamefont {{Monkhors H.}}}\ and\
  \bibinfo {author} {\bibfnamefont {J.}~\bibnamefont {Pack}},\ }\href {\doibase
  10.1103/PhysRevB.13.5188} {\bibfield  {journal} {\bibinfo  {journal} {Phys.
  Rev. B}\ }\textbf {\bibinfo {volume} {13}},\ \bibinfo {pages} {5188}
  (\bibinfo {year} {1976})}\BibitemShut {NoStop}%
\bibitem [{\citenamefont {Methfessel}\ and\ \citenamefont
  {Paxton}(1989)}]{Methfessel1989}%
  \BibitemOpen
  \bibfield  {author} {\bibinfo {author} {\bibfnamefont {M.}~\bibnamefont
  {Methfessel}}\ and\ \bibinfo {author} {\bibfnamefont {A.~T.}\ \bibnamefont
  {Paxton}},\ }\href {\doibase 10.1103/PhysRevB.40.3616} {\bibfield  {journal}
  {\bibinfo  {journal} {Phys. Rev. B}\ }\textbf {\bibinfo {volume} {40}},\
  \bibinfo {pages} {3616} (\bibinfo {year} {1989})}\BibitemShut {NoStop}%
\bibitem [{\citenamefont {Bl{\"{o}}chl}\ \emph {et~al.}(1994)\citenamefont
  {Bl{\"{o}}chl}, \citenamefont {Jepsen},\ and\ \citenamefont
  {Andersen}}]{Blochl1994}%
  \BibitemOpen
  \bibfield  {author} {\bibinfo {author} {\bibfnamefont {P.~E.}\ \bibnamefont
  {Bl{\"{o}}chl}}, \bibinfo {author} {\bibfnamefont {O.}~\bibnamefont
  {Jepsen}}, \ and\ \bibinfo {author} {\bibfnamefont {O.~K.}\ \bibnamefont
  {Andersen}},\ }\href {\doibase 10.1103/PhysRevB.49.16223} {\bibfield
  {journal} {\bibinfo  {journal} {Phys. Rev. B}\ }\textbf {\bibinfo {volume}
  {49}},\ \bibinfo {pages} {16223} (\bibinfo {year} {1994})}\BibitemShut
  {NoStop}%
\bibitem [{\citenamefont {Deringer}\ \emph {et~al.}(2011)\citenamefont
  {Deringer}, \citenamefont {Tchougr{\'{e}}eff},\ and\ \citenamefont
  {Dronskowski}}]{Deringer2011}%
  \BibitemOpen
  \bibfield  {author} {\bibinfo {author} {\bibfnamefont {V.~L.}\ \bibnamefont
  {Deringer}}, \bibinfo {author} {\bibfnamefont {A.~L.}\ \bibnamefont
  {Tchougr{\'{e}}eff}}, \ and\ \bibinfo {author} {\bibfnamefont
  {R.}~\bibnamefont {Dronskowski}},\ }\href {\doibase 10.1021/jp202489s}
  {\bibfield  {journal} {\bibinfo  {journal} {J. Phys. Chem. A}\ }\textbf
  {\bibinfo {volume} {115}},\ \bibinfo {pages} {5461} (\bibinfo {year}
  {2011})}\BibitemShut {NoStop}%
\bibitem [{\citenamefont {Maintz}\ \emph {et~al.}(2013)\citenamefont {Maintz},
  \citenamefont {Deringer}, \citenamefont {Tchougr{\'{e}}eff},\ and\
  \citenamefont {Dronskowski}}]{Maintz2013}%
  \BibitemOpen
  \bibfield  {author} {\bibinfo {author} {\bibfnamefont {S.}~\bibnamefont
  {Maintz}}, \bibinfo {author} {\bibfnamefont {V.~L.}\ \bibnamefont
  {Deringer}}, \bibinfo {author} {\bibfnamefont {A.~L.}\ \bibnamefont
  {Tchougr{\'{e}}eff}}, \ and\ \bibinfo {author} {\bibfnamefont
  {R.}~\bibnamefont {Dronskowski}},\ }\href {\doibase 10.1002/jcc.23424}
  {\bibfield  {journal} {\bibinfo  {journal} {J. Comput. Chem.}\ }\textbf
  {\bibinfo {volume} {34}},\ \bibinfo {pages} {2557} (\bibinfo {year}
  {2013})}\BibitemShut {NoStop}%
\bibitem [{\citenamefont {Maintz}\ \emph
  {et~al.}(2016{\natexlab{a}})\citenamefont {Maintz}, \citenamefont {Esser},\
  and\ \citenamefont {Dronskowski}}]{Maintz2016a}%
  \BibitemOpen
  \bibfield  {author} {\bibinfo {author} {\bibfnamefont {S.}~\bibnamefont
  {Maintz}}, \bibinfo {author} {\bibfnamefont {M.}~\bibnamefont {Esser}}, \
  and\ \bibinfo {author} {\bibfnamefont {R.}~\bibnamefont {Dronskowski}},\
  }\href {\doibase 10.5506/APhysPolB.47.1165} {\bibfield  {journal} {\bibinfo
  {journal} {Acta Phys. Pol. B}\ }\textbf {\bibinfo {volume} {47}},\ \bibinfo
  {pages} {1165} (\bibinfo {year} {2016}{\natexlab{a}})}\BibitemShut {NoStop}%
\bibitem [{\citenamefont {Dronskowski}\ and\ \citenamefont
  {Bl{\"{o}}chl}(1993)}]{Dronskowski1993}%
  \BibitemOpen
  \bibfield  {author} {\bibinfo {author} {\bibfnamefont {R.}~\bibnamefont
  {Dronskowski}}\ and\ \bibinfo {author} {\bibfnamefont {P.~E.}\ \bibnamefont
  {Bl{\"{o}}chl}},\ }\href {\doibase 10.1021/j100135a014} {\bibfield  {journal}
  {\bibinfo  {journal} {J. Phys. Chem.}\ }\textbf {\bibinfo {volume} {97}},\
  \bibinfo {pages} {8617} (\bibinfo {year} {1993})}\BibitemShut {NoStop}%
\bibitem [{\citenamefont {Maintz}\ \emph
  {et~al.}(2016{\natexlab{b}})\citenamefont {Maintz}, \citenamefont {Deringer},
  \citenamefont {Tchougr{\'{e}}eff},\ and\ \citenamefont
  {Dronskowski}}]{Maintz2016}%
  \BibitemOpen
  \bibfield  {author} {\bibinfo {author} {\bibfnamefont {S.}~\bibnamefont
  {Maintz}}, \bibinfo {author} {\bibfnamefont {V.~L.}\ \bibnamefont
  {Deringer}}, \bibinfo {author} {\bibfnamefont {A.~L.}\ \bibnamefont
  {Tchougr{\'{e}}eff}}, \ and\ \bibinfo {author} {\bibfnamefont
  {R.}~\bibnamefont {Dronskowski}},\ }\href {\doibase 10.1002/jcc.24300}
  {\bibfield  {journal} {\bibinfo  {journal} {J. Comput. Chem.}\ }\textbf
  {\bibinfo {volume} {37}},\ \bibinfo {pages} {1030} (\bibinfo {year}
  {2016}{\natexlab{b}})}\BibitemShut {NoStop}%
\bibitem [{\citenamefont {Thomas}\ and\ \citenamefont {{Van der
  Ven}}(2013)}]{Thomas2013}%
  \BibitemOpen
  \bibfield  {author} {\bibinfo {author} {\bibfnamefont {J.~C.}\ \bibnamefont
  {Thomas}}\ and\ \bibinfo {author} {\bibfnamefont {A.}~\bibnamefont {{Van der
  Ven}}},\ }\href {\doibase 10.1103/PhysRevB.88.214111} {\bibfield  {journal}
  {\bibinfo  {journal} {Phys. Rev. B}\ }\textbf {\bibinfo {volume} {88}},\
  \bibinfo {pages} {214111} (\bibinfo {year} {2013})}\BibitemShut {NoStop}%
\bibitem [{\citenamefont {Puchala}\ \emph {et~al.}(2016)\citenamefont
  {Puchala}, \citenamefont {Radin}, \citenamefont {Gunda}, \citenamefont
  {Goiri}, \citenamefont {Decolvenaere}, \citenamefont {Natarajan},\ and\
  \citenamefont {Thomas}}]{Puchala2016}%
  \BibitemOpen
  \bibfield  {author} {\bibinfo {author} {\bibfnamefont {B.}~\bibnamefont
  {Puchala}}, \bibinfo {author} {\bibfnamefont {M.}~\bibnamefont {Radin}},
  \bibinfo {author} {\bibfnamefont {N.~S.~H.}\ \bibnamefont {Gunda}}, \bibinfo
  {author} {\bibfnamefont {J.}~\bibnamefont {Goiri}}, \bibinfo {author}
  {\bibfnamefont {E.}~\bibnamefont {Decolvenaere}}, \bibinfo {author}
  {\bibfnamefont {A.~R.}\ \bibnamefont {Natarajan}}, \ and\ \bibinfo {author}
  {\bibfnamefont {J.~C.}\ \bibnamefont {Thomas}},\ }\href {\doibase
  10.5281/zenodo.546148} {\  (\bibinfo {year} {2016}),\
  10.5281/zenodo.546148}\BibitemShut {NoStop}%
\bibitem [{\citenamefont {Puchala}\ and\ \citenamefont {{Van der
  Ven}}(2013)}]{Puchala2013}%
  \BibitemOpen
  \bibfield  {author} {\bibinfo {author} {\bibfnamefont {B.}~\bibnamefont
  {Puchala}}\ and\ \bibinfo {author} {\bibfnamefont {A.}~\bibnamefont {{Van der
  Ven}}},\ }\href {\doibase 10.1103/PhysRevB.88.094108} {\bibfield  {journal}
  {\bibinfo  {journal} {Phys. Rev. B}\ }\textbf {\bibinfo {volume} {88}},\
  \bibinfo {pages} {094108} (\bibinfo {year} {2013})}\BibitemShut {NoStop}%
\bibitem [{\citenamefont {{Van der Ven}}\ \emph {et~al.}(2010)\citenamefont
  {{Van der Ven}}, \citenamefont {Thomas}, \citenamefont {Xu},\ and\
  \citenamefont {Bhattacharya}}]{VanderVen2010}%
  \BibitemOpen
  \bibfield  {author} {\bibinfo {author} {\bibfnamefont {A.}~\bibnamefont {{Van
  der Ven}}}, \bibinfo {author} {\bibfnamefont {J.~C.}\ \bibnamefont {Thomas}},
  \bibinfo {author} {\bibfnamefont {Q.}~\bibnamefont {Xu}}, \ and\ \bibinfo
  {author} {\bibfnamefont {J.}~\bibnamefont {Bhattacharya}},\ }\href {\doibase
  10.1016/j.matcom.2009.08.008} {\bibfield  {journal} {\bibinfo  {journal}
  {Math. Comput. Simul.}\ }\textbf {\bibinfo {volume} {80}},\ \bibinfo {pages}
  {1393} (\bibinfo {year} {2010})}\BibitemShut {NoStop}%
\bibitem [{\citenamefont {Schimka}\ \emph {et~al.}(2013)\citenamefont
  {Schimka}, \citenamefont {Gaudoin}, \citenamefont {Klime{\v{s}}},
  \citenamefont {Marsman},\ and\ \citenamefont {Kresse}}]{Schimka2013}%
  \BibitemOpen
  \bibfield  {author} {\bibinfo {author} {\bibfnamefont {L.}~\bibnamefont
  {Schimka}}, \bibinfo {author} {\bibfnamefont {R.}~\bibnamefont {Gaudoin}},
  \bibinfo {author} {\bibfnamefont {J.}~\bibnamefont {Klime{\v{s}}}}, \bibinfo
  {author} {\bibfnamefont {M.}~\bibnamefont {Marsman}}, \ and\ \bibinfo
  {author} {\bibfnamefont {G.}~\bibnamefont {Kresse}},\ }\href {\doibase
  10.1103/PhysRevB.87.214102} {\bibfield  {journal} {\bibinfo  {journal} {Phys.
  Rev. B}\ }\textbf {\bibinfo {volume} {87}},\ \bibinfo {pages} {214102}
  (\bibinfo {year} {2013})}\BibitemShut {NoStop}%
\bibitem [{Note4()}]{Note4}%
  \BibitemOpen
  \bibinfo {note} {Non-spin-polarized should not be taken to mean paramagnetic,
  in this context. While non-spin-polarized calculations have sometimes been
  used as a proxy for paramagnetic configurations, we stress that this approach
  is generally incorrect.}\BibitemShut {Stop}%
\bibitem [{\citenamefont {Thoene}\ \emph {et~al.}(2009)\citenamefont {Thoene},
  \citenamefont {Chadov}, \citenamefont {Fecher}, \citenamefont {Felser},\ and\
  \citenamefont {K{\"{u}}bler}}]{Thoene2009}%
  \BibitemOpen
  \bibfield  {author} {\bibinfo {author} {\bibfnamefont {J.}~\bibnamefont
  {Thoene}}, \bibinfo {author} {\bibfnamefont {S.}~\bibnamefont {Chadov}},
  \bibinfo {author} {\bibfnamefont {G.}~\bibnamefont {Fecher}}, \bibinfo
  {author} {\bibfnamefont {C.}~\bibnamefont {Felser}}, \ and\ \bibinfo {author}
  {\bibfnamefont {J.}~\bibnamefont {K{\"{u}}bler}},\ }\href {\doibase
  10.1088/0022-3727/42/8/084013} {\bibfield  {journal} {\bibinfo  {journal} {J.
  Phys. D. Appl. Phys.}\ }\textbf {\bibinfo {volume} {42}},\ \bibinfo {pages}
  {084013} (\bibinfo {year} {2009})}\BibitemShut {NoStop}%
\bibitem [{\citenamefont {Gr{\"{u}}nebohm}\ \emph {et~al.}(2016)\citenamefont
  {Gr{\"{u}}nebohm}, \citenamefont {Herper},\ and\ \citenamefont
  {Entel}}]{Grunebohm2016}%
  \BibitemOpen
  \bibfield  {author} {\bibinfo {author} {\bibfnamefont {A.}~\bibnamefont
  {Gr{\"{u}}nebohm}}, \bibinfo {author} {\bibfnamefont {H.~C.}\ \bibnamefont
  {Herper}}, \ and\ \bibinfo {author} {\bibfnamefont {P.}~\bibnamefont
  {Entel}},\ }\href {\doibase 10.1088/0022-3727/49/39/395001} {\bibfield
  {journal} {\bibinfo  {journal} {J. Phys. D. Appl. Phys.}\ }\textbf {\bibinfo
  {volume} {49}},\ \bibinfo {pages} {395001} (\bibinfo {year} {2016})},\
  \Eprint {http://arxiv.org/abs/1609.09399} {arXiv:1609.09399} \BibitemShut
  {NoStop}%
\bibitem [{\citenamefont {Landrum}\ and\ \citenamefont
  {Dronskowski}(2000)}]{Landrum2000}%
  \BibitemOpen
  \bibfield  {author} {\bibinfo {author} {\bibfnamefont {G.}~\bibnamefont
  {Landrum}}\ and\ \bibinfo {author} {\bibfnamefont {R.}~\bibnamefont
  {Dronskowski}},\ }\href {\doibase
  10.1002/(SICI)1521-3773(20000502)39:9<1560::AID-ANIE1560>3.0.CO;2-T}
  {\bibfield  {journal} {\bibinfo  {journal} {Angew. Chem. Int. Ed. Engl.}\
  }\textbf {\bibinfo {volume} {39}},\ \bibinfo {pages} {1560} (\bibinfo {year}
  {2000})}\BibitemShut {NoStop}%
\bibitem [{Note5()}]{Note5}%
  \BibitemOpen
  \bibinfo {note} {It is worth noting that, had we enumerated larger
  supercells, the lowest-energy structures in the intermediate composition
  range would asymptotically approach zero. This is a result of the Mn/Fe
  miscibility gap: larger structures would allow larger and larger volumes of
  Mn-rich and Fe-rich domains, decreasing the surface-area-to-volume ratio of
  the interface.}\BibitemShut {Stop}%
\bibitem [{\citenamefont {Ishida}\ \emph
  {et~al.}(1995{\natexlab{b}})\citenamefont {Ishida}, \citenamefont {Fujii},
  \citenamefont {Kashiwagi},\ and\ \citenamefont {Asano}}]{Ishida1995}%
  \BibitemOpen
  \bibfield  {author} {\bibinfo {author} {\bibfnamefont {S.}~\bibnamefont
  {Ishida}}, \bibinfo {author} {\bibfnamefont {S.}~\bibnamefont {Fujii}},
  \bibinfo {author} {\bibfnamefont {S.}~\bibnamefont {Kashiwagi}}, \ and\
  \bibinfo {author} {\bibfnamefont {S.}~\bibnamefont {Asano}},\ }\href
  {\doibase 10.1143/JPSJ.64.2152} {\bibfield  {journal} {\bibinfo  {journal}
  {J. Phys. Soc. Japan}\ }\textbf {\bibinfo {volume} {64}},\ \bibinfo {pages}
  {2152} (\bibinfo {year} {1995}{\natexlab{b}})}\BibitemShut {NoStop}%
\bibitem [{\citenamefont {Galanakis}\ and\ \citenamefont
  {Mavropoulos}(2007)}]{Galanakis2007}%
  \BibitemOpen
  \bibfield  {author} {\bibinfo {author} {\bibfnamefont {I.}~\bibnamefont
  {Galanakis}}\ and\ \bibinfo {author} {\bibfnamefont {P.}~\bibnamefont
  {Mavropoulos}},\ }\href {\doibase 10.1088/0953-8984/19/31/315213} {\bibfield
  {journal} {\bibinfo  {journal} {J. Phys. Condens. Matter}\ }\textbf {\bibinfo
  {volume} {19}},\ \bibinfo {pages} {315213} (\bibinfo {year} {2007})},\
  \Eprint {http://arxiv.org/abs/0610827} {arXiv:0610827 [cond-mat]}
  \BibitemShut {NoStop}%
\bibitem [{Note6()}]{Note6}%
  \BibitemOpen
  \bibinfo {note} {While the choice of functional (\protect \textit {e.g.},
  GGA-PBE vs. LDA) is known to influence the \protect \textit {size} of the
  gap, the choice of functional should not impact whether a gap is observed at
  all.\cite {Rai2016}}\BibitemShut {NoStop}%
\bibitem [{\citenamefont {N{\'{e}}el}(1936)}]{Neel1936}%
  \BibitemOpen
  \bibfield  {author} {\bibinfo {author} {\bibfnamefont {L.}~\bibnamefont
  {N{\'{e}}el}},\ }\href@noop {} {\bibfield  {journal} {\bibinfo  {journal}
  {Comptes Rendus l'Acad{\'{e}}mie des Sci.}\ }\textbf {\bibinfo {volume}
  {203}},\ \bibinfo {pages} {304} (\bibinfo {year} {1936})}\BibitemShut
  {NoStop}%
\bibitem [{\citenamefont {Rai}\ \emph {et~al.}(2016)\citenamefont {Rai},
  \citenamefont {Sandeep}, \citenamefont {Shankar}, \citenamefont {Sakhya},
  \citenamefont {Sinha}, \citenamefont {Khenata}, \citenamefont {Ghimire},\
  and\ \citenamefont {Thapa}}]{Rai2016}%
  \BibitemOpen
  \bibfield  {author} {\bibinfo {author} {\bibfnamefont {D.~P.}\ \bibnamefont
  {Rai}}, \bibinfo {author} {\bibnamefont {Sandeep}}, \bibinfo {author}
  {\bibfnamefont {A.}~\bibnamefont {Shankar}}, \bibinfo {author} {\bibfnamefont
  {A.~P.}\ \bibnamefont {Sakhya}}, \bibinfo {author} {\bibfnamefont {T.~P.}\
  \bibnamefont {Sinha}}, \bibinfo {author} {\bibfnamefont {R.}~\bibnamefont
  {Khenata}}, \bibinfo {author} {\bibfnamefont {M.~P.}\ \bibnamefont
  {Ghimire}}, \ and\ \bibinfo {author} {\bibfnamefont {R.~K.}\ \bibnamefont
  {Thapa}},\ }\href {\doibase 10.1088/2053-1591/3/7/075022} {\bibfield
  {journal} {\bibinfo  {journal} {Mater. Res. Express}\ }\textbf {\bibinfo
  {volume} {3}},\ \bibinfo {pages} {075022} (\bibinfo {year}
  {2016})}\BibitemShut {NoStop}%
\end{thebibliography}%

\end{document}